\documentclass[twocolumn]{aastex62}
\usepackage{IEEEtrantools}
\bstctlcite{IEEEexample:BSTcontrol}
\usepackage{natbib}
\usepackage{amsmath}
\usepackage{mathtools}
\usepackage{rotating}
\usepackage{lipsum}

\usepackage{multirow}
\usepackage{makecell} 

\received{June 7, 2020}
\revised{August 1, 2020}
\accepted{August 13, 2020}

\shorttitle{Ghostly DLAs in SDSS DR14}
\shortauthors{Hassan Fathivavsari}

\begin{document}

\title{\bf{Using Machine Learning to Find Ghostly Damped Ly$\alpha$ Systems in SDSS DR14}}

\correspondingauthor{Hassan Fathivavsari}
\email{h.fathie@gmail.com}

\author{Hassan Fathivavsari}
\affiliation{School of Astronomy, Institute for Research in Fundamental Sciences (IPM), P. O. Box 19395-5531, Tehran, Iran}

\begin{abstract}
We report the discovery of 59 new ghostly absorbers from the Sloan Digital Sky Survey (SDSS) Data Release 14 (DR14). These absorbers, with $z_{\rm abs}$\,$\sim$\,$z_{\rm QSO}$, reveal no Ly$\alpha$ absorption, and they are mainly identified through the detection of strong metal absorption lines  in the spectra. The number of previously known such systems is 30. The new systems are found with the aid of machine learning algorithms. The spectra of 41 (out of total of 89) absorbers also cover the Ly$\beta$ spectral region. By fitting the damping wings of the Ly$\beta$ absorption in the stacked spectrum of 21 (out of 41) absorbers with relatively stronger Ly$\beta$ absorption, we measured an H\,{\sc i} column density of log\,$N$(H\,{\sc i})\,=\,21.50. This column density is 0.5\,dex higher than that of the previous work. We also found that the metal absorption lines in the stacked spectrum of the 21 ghostly absorbers with stronger Ly$\beta$ absorption have similar properties as those in the stacked spectrum of the remaining systems. These circumstantial evidence strongly suggest that the majority of our ghostly absorbers are indeed DLAs.

\end{abstract}


\section{Introduction} \label{sec:intro}

\defcitealias{2020ApJ...888...85F}{Paper\,I}

Neutral hydrogen gas clouds with H\,{\sc i} column density log\,$N$(H\,{\sc i})\,$\ge$\,20.3, located along the line of sight to distant quasars, are recognized by the damped Ly$\alpha$ (DLA) absorption lines they produce in the spectra of their background quasars \citep{1989ApJ...344..567T,2005ARA&A..43..861W}. These so-called DLA absorbers are classified into two main categories: intervening and associated. Intervening DLA absorbers are neutral hydrogen gas clouds that happen to be located along the line of sight to the background quasars, but still unrelated to the quasars themselves. Detailed studies of these absorbers indicate an origin from the interstellar medium (ISM) of galaxies or circumgalactic medium \citep{1997ApJ...487...73P,1998ApJ...507..113P,1998ApJ...495..647H,2000ApJ...534..594H,2015MNRAS.447.1834B}. Recent detection of galaxies in emission at the same redshift of intervening DLAs further confirms the connection between these absorbers and the ISM of galaxies \citep{2018ApJ...856L..23K,2018MNRAS.474.4039M,2018ApJ...856L..12N,2019ApJ...870L..19N,2019MNRAS.482L..65K}. On the other hand, associated DLAs originate from gas clouds that are physically close to their background quasars, and are also usually affected by the quasars through radiation or mechanical feedback. These absorbers may arise from infalling and/or outflowing gas, the ISM of the quasar host galaxy, or the ISM of the neighboring galaxies in the group environment \citep{2002A&A...383...91E,2010MNRAS.406.1435E,2011MNRAS.412..448E}.

The distinguishing between intervening and associated DLAs is not observationally straightforward. However, by convention, absorbers with the apparent velocity difference $\Delta$$V$\,$\ge$\,3000\,km\,s$^{-1}$ are usually assumed to be intervening \citep{1991ApJS...77....1L,1996ApJ...468..121S,2006ApJ...646..730J,2009ApJ...696.1543P}. This velocity cut is motivated by the argument that the peculiar velocities of quasars outflowing gas are less likely to reach such high values. The situation for DLA absorbers with $\Delta$$V$\,$<$\,3000\,km\,s$^{-1}$ is more complicated as these absorbers could be either associated (e.g. high-velocity ejected material) or intervening. Due to this complexity, these DLAs are usually called proximate-DLAs (PDLAs) until further information about their origin is available. 

When a PDLA is located at the redshift of the quasar, it can act as a natural coronagraph, blocking the Ly$\alpha$ emission from the broad line region (BLR) of the quasar \citep{2009ApJ...693L..49H,2013A&A...558A.111F}. Since these PDLAs eclipse their background quasars in Ly$\alpha$, they are also called \emph{eclipsing} DLAs \citep{2015MNRAS.454..876F,2016MNRAS.461.1816F,2018MNRAS.477.5625F}. If the projected size of an eclipsing DLA is smaller than the narrow line region (NLR) and/or star-forming regions in the host galaxy, then the Ly$\alpha$ emission from these regions could be detected as some narrow Ly$\alpha$ emission line in the Ly$\alpha$ absorption core of the eclipsing DLA. \citet{2018MNRAS.477.5625F} recently searched the SDSS-III DR12 spectra and found 399 eclipsing DLAs with log\,$N$(H\,{\sc i})\,$\ge$\,21.10. By analyzing the stacked spectra of these eclipsing DLAs, they found that absorption from high ionization species (i.e. N\,{\sc v}, Si\,{\sc iv}, and C\,{\sc iv}) and also from the fine structure states of Si\,{\sc ii} and C\,{\sc ii} are stronger in eclipsing DLAs that exhibit stronger Ly$\alpha$ emission in their Ly$\alpha$ absorption troughs.

In some extreme cases in which an eclipsing DLA is much smaller than the BLR, the leaked Ly$\alpha$ emission from the regions of the BLR that are not covered by the DLA, could fill the DLA absorption trough, and consequently form a \emph{ghostly} DLA in the quasar spectrum \citep{2016ApJ...821....1J,2017MNRAS.466L..58F,2020ApJ...888...85F,2020MNRAS.495..460D}. These DLAs are dubbed ghostly because their DLA absorption lines are not visible in the spectra. Recently, \citet{2020ApJ...888...85F} found 30 ghostly absorbers in the SDSS-III DR12 spectra. They derived an H\,{\sc i} column density of log\,$N$(H\,{\sc i})\,$\sim$\,21.0 by fitting the Ly$\beta$ absorption in the stacked spectrum of 13 absorbers whose Ly$\beta$ spectral regions were covered in the observed spectral window. By comparing the properties of ghostly absorbers with those of eclipsing DLAs, they discovered a sequence in the observed properties of these absorbers with ghostly absorbers exhibiting broader H\,{\sc i} kinematics, stronger absorption from high ionization species, stronger absorption from the excited states of C\,{\sc ii} and Si\,{\sc ii}, and a higher level of dust extinction. From these observations, they conclude that ghostly absorbers are from the same population as eclipsing DLAs, except that ghostly absorbers are denser and located closer to the central quasar.

In the current paper, we use machine learning to search for and find ghostly DLAs in the SDSS DR14 spectra. By compiling a larger sample of ghostly absorbers we revisit the results from \citet[][hereafter Paper\,I]{2020ApJ...888...85F}.

\begin{figure*}
\centering
\begin{tabular}{c}
\includegraphics[width=0.99\hsize]{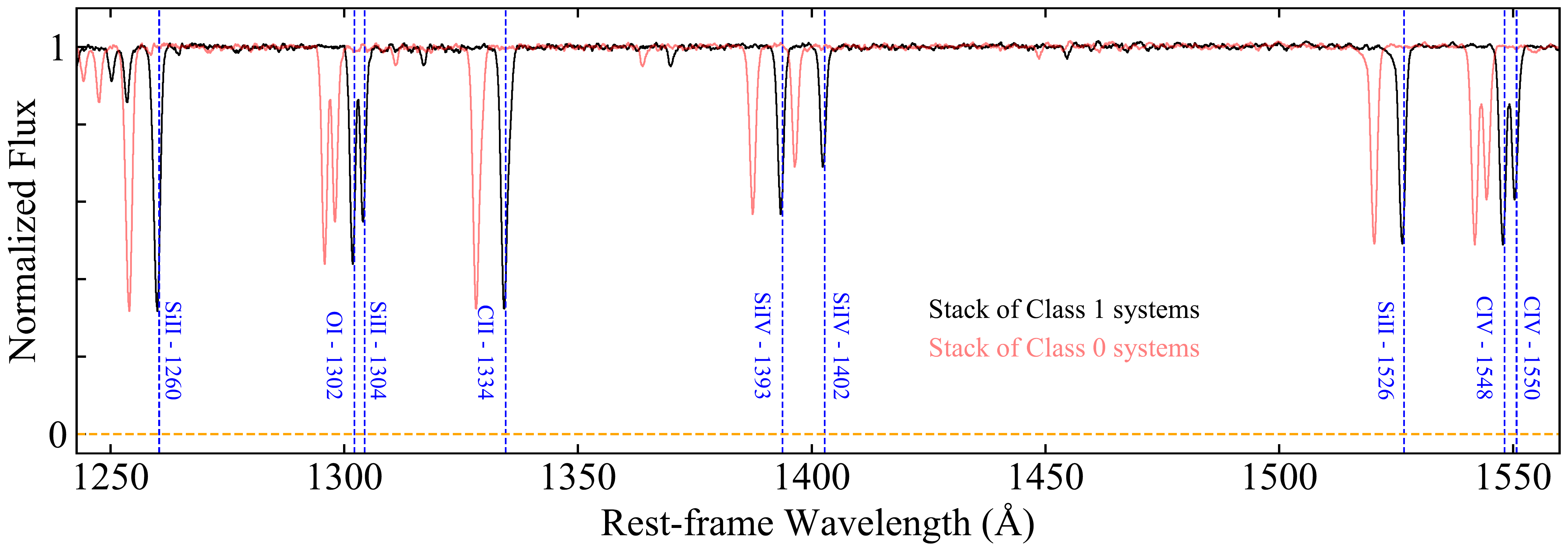}
\end{tabular}
\caption{The median-stacked spectra of the class\,1 (the black spectrum) and class\,0 (the red spectrum) systems. As expected, no absorption lines are seen at the expected position of the absorption lines in the class\,0 stacked spectrum}
 \label{class_0_1}
\end{figure*}

\begin{table}
\caption{XGBoost hyperparameters before and after parameter optimization.}
\centering 
\setlength{\tabcolsep}{6.8pt}
\renewcommand{\arraystretch}{1.05}
\begin{tabular}{c c c} 
\hline\hline
Hyperparameters & Default values & Optimal values  \\
\hline 
\multicolumn{1}{l}{\texttt{n\_estimators}} & 100 & 400 \\
\multicolumn{1}{l}{\texttt{max\_depth}} & 6 & 4 \\
\multicolumn{1}{l}{\texttt{learning\_rate}} & 0.3 & 0.04 \\
\multicolumn{1}{l}{\texttt{min\_child\_weight}} & 1 & 1 \\
\multicolumn{1}{l}{\texttt{gamma}} & 0 & 0.2 \\
\multicolumn{1}{l}{\texttt{subsample}} & 1 & 0.5 \\
\multicolumn{1}{l}{\texttt{colsample\_bytree}} & 1 & 0.2 \\
\hline
\end{tabular}
\label{table1}
\end{table}

\begin{figure*}
\centering
\begin{tabular}{c}
\includegraphics[width=0.99\hsize]{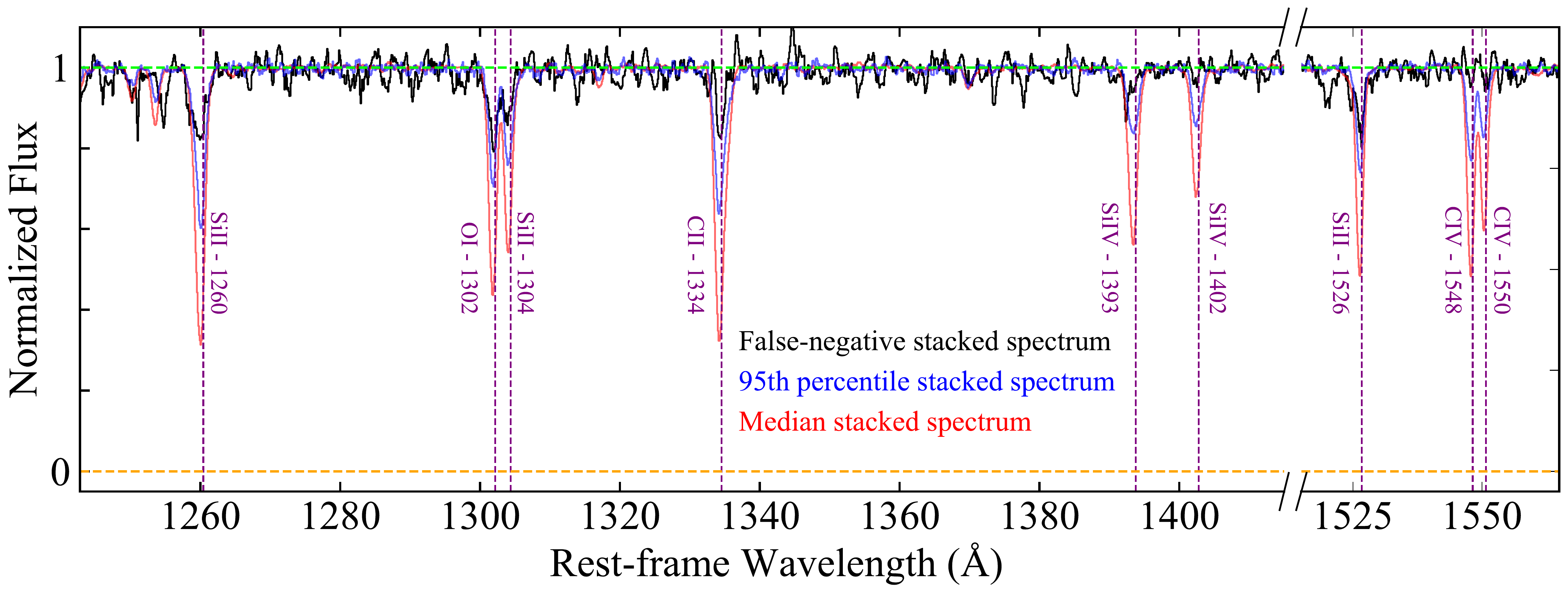}
\end{tabular}
\caption{The median-stacked spectrum of the 9 false negative systems (the black spectrum), the median-stacked spectrum of the 641 true positive systems (the red spectrum), and the 95th percentile stacked spectrum of the true positive systems (the blue spectrum).}
 \label{FNStack}
\end{figure*}

\begin{figure*}
\centering
\begin{tabular}{cccc}
\includegraphics[width=0.23\hsize]{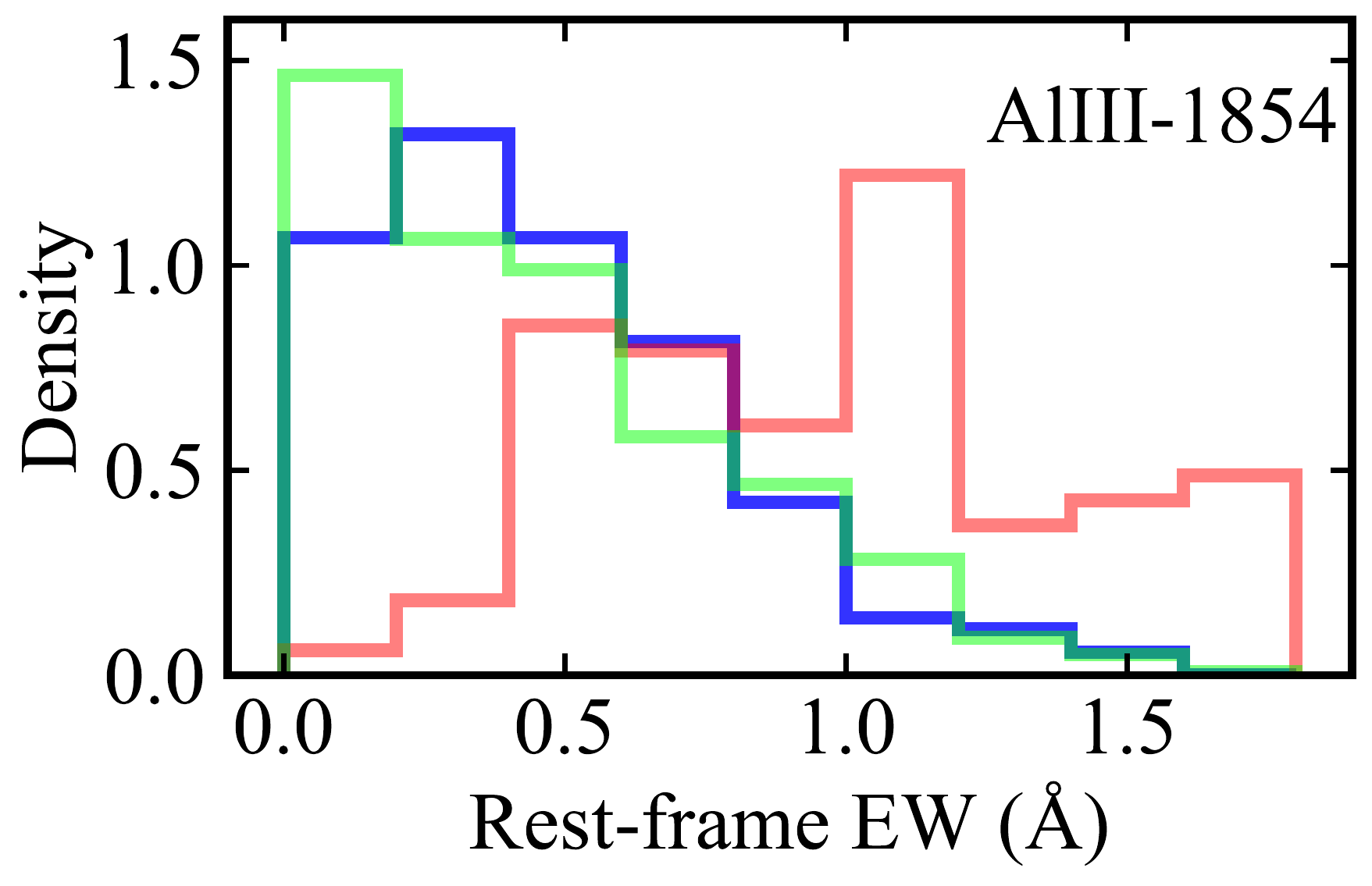}
\includegraphics[width=0.23\hsize]{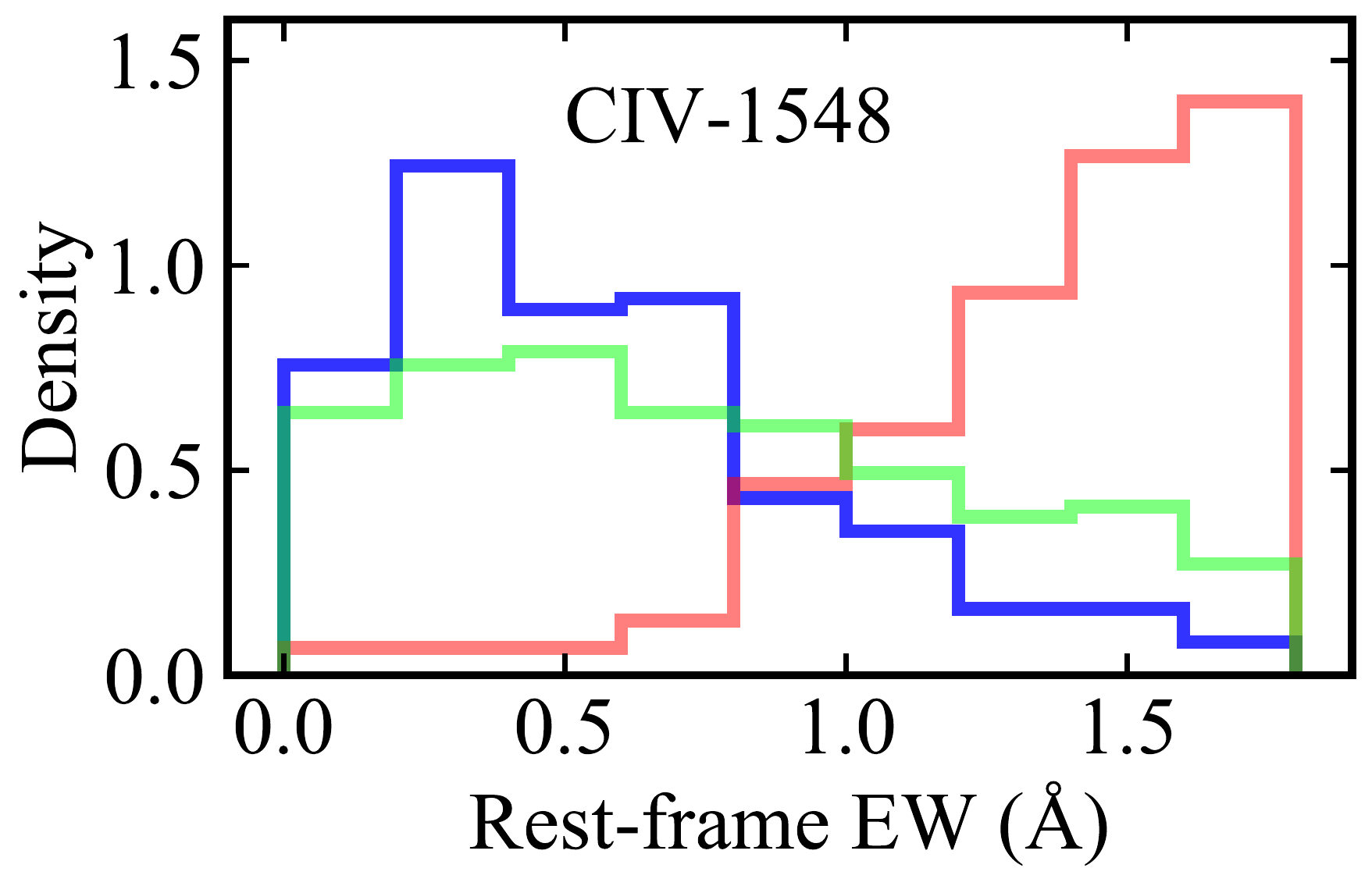}
\includegraphics[width=0.23\hsize]{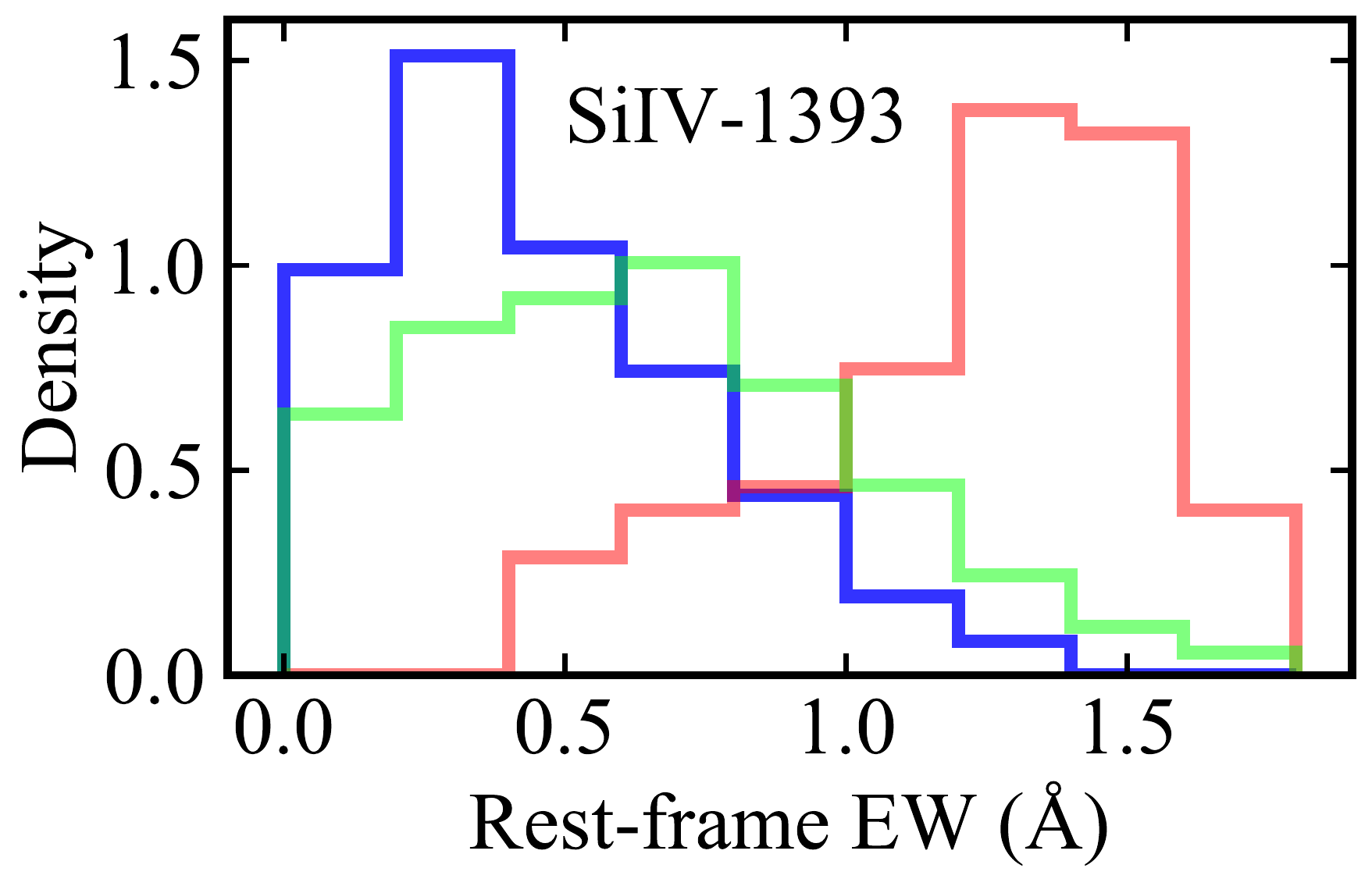}
\includegraphics[width=0.23\hsize]{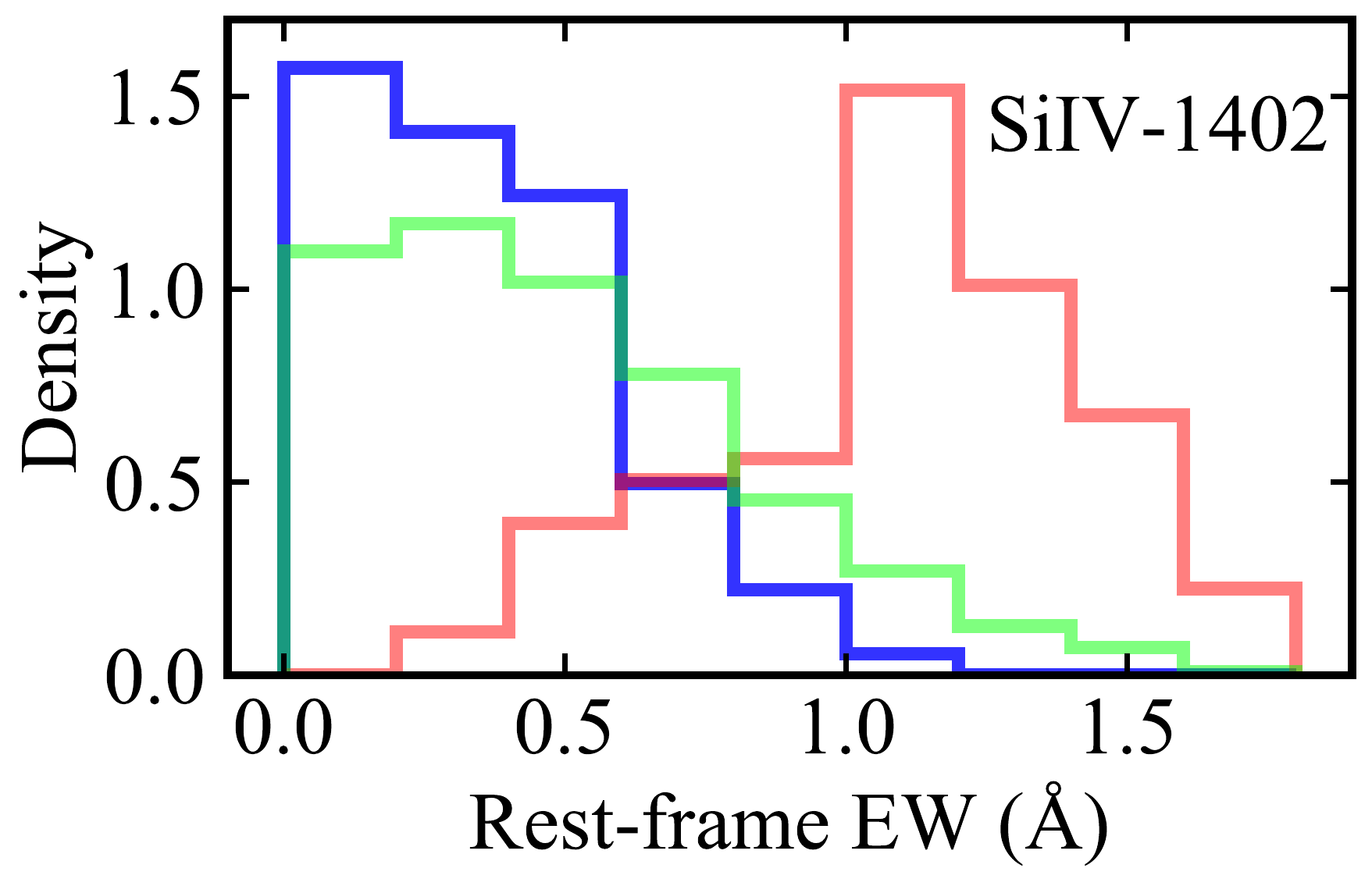} \\
\includegraphics[width=0.23\hsize]{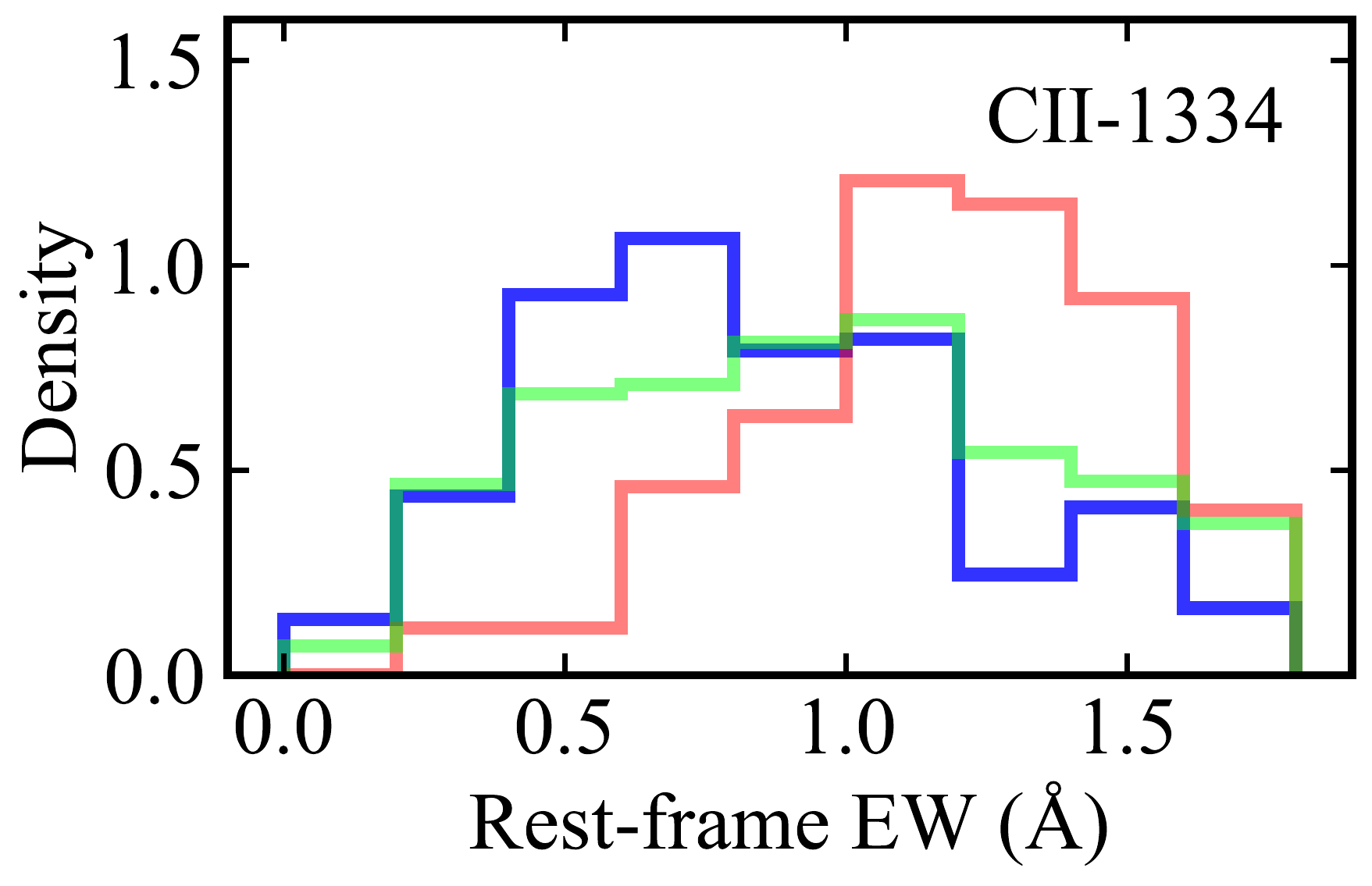}
\includegraphics[width=0.23\hsize]{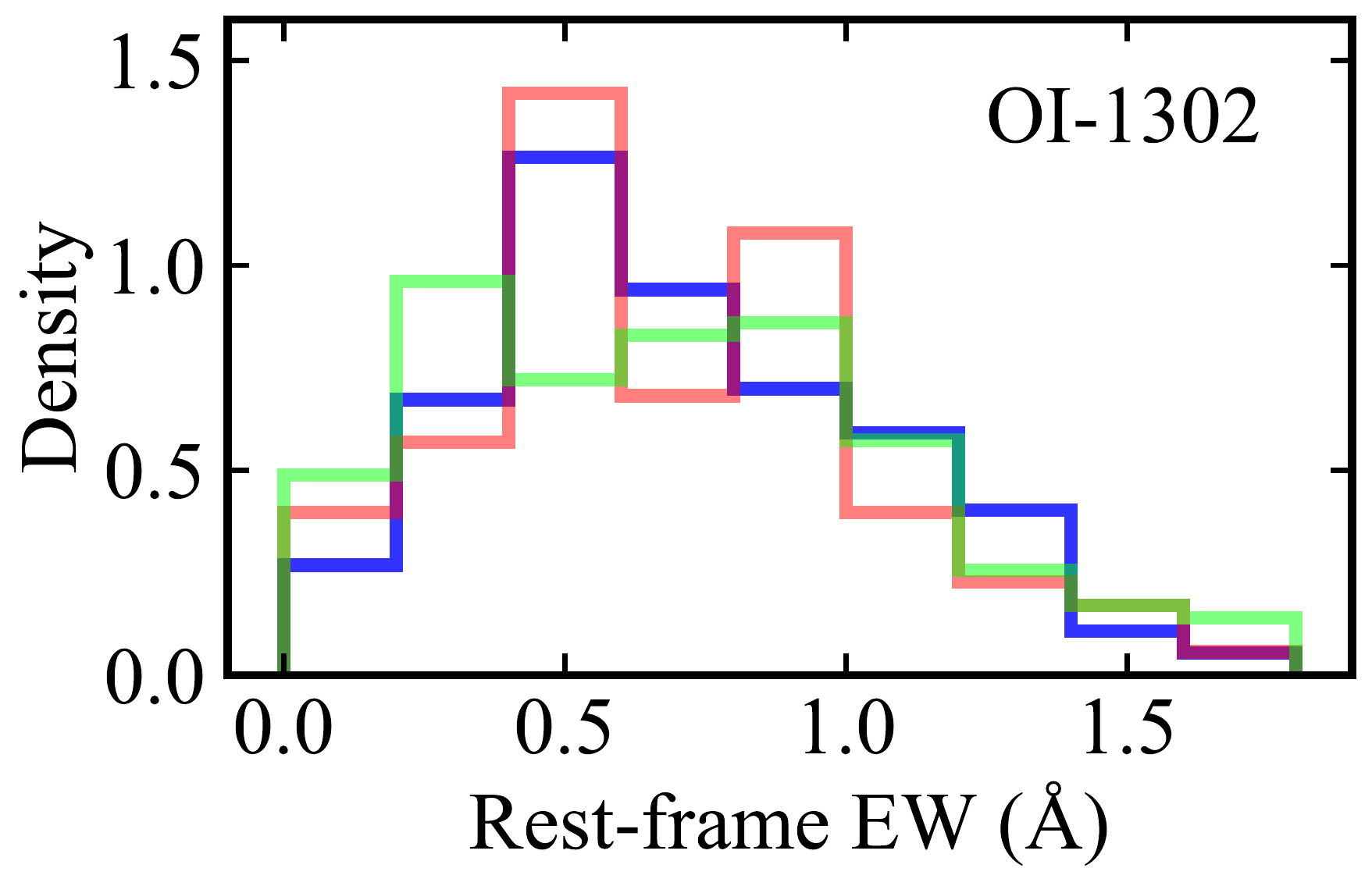}
\includegraphics[width=0.23\hsize]{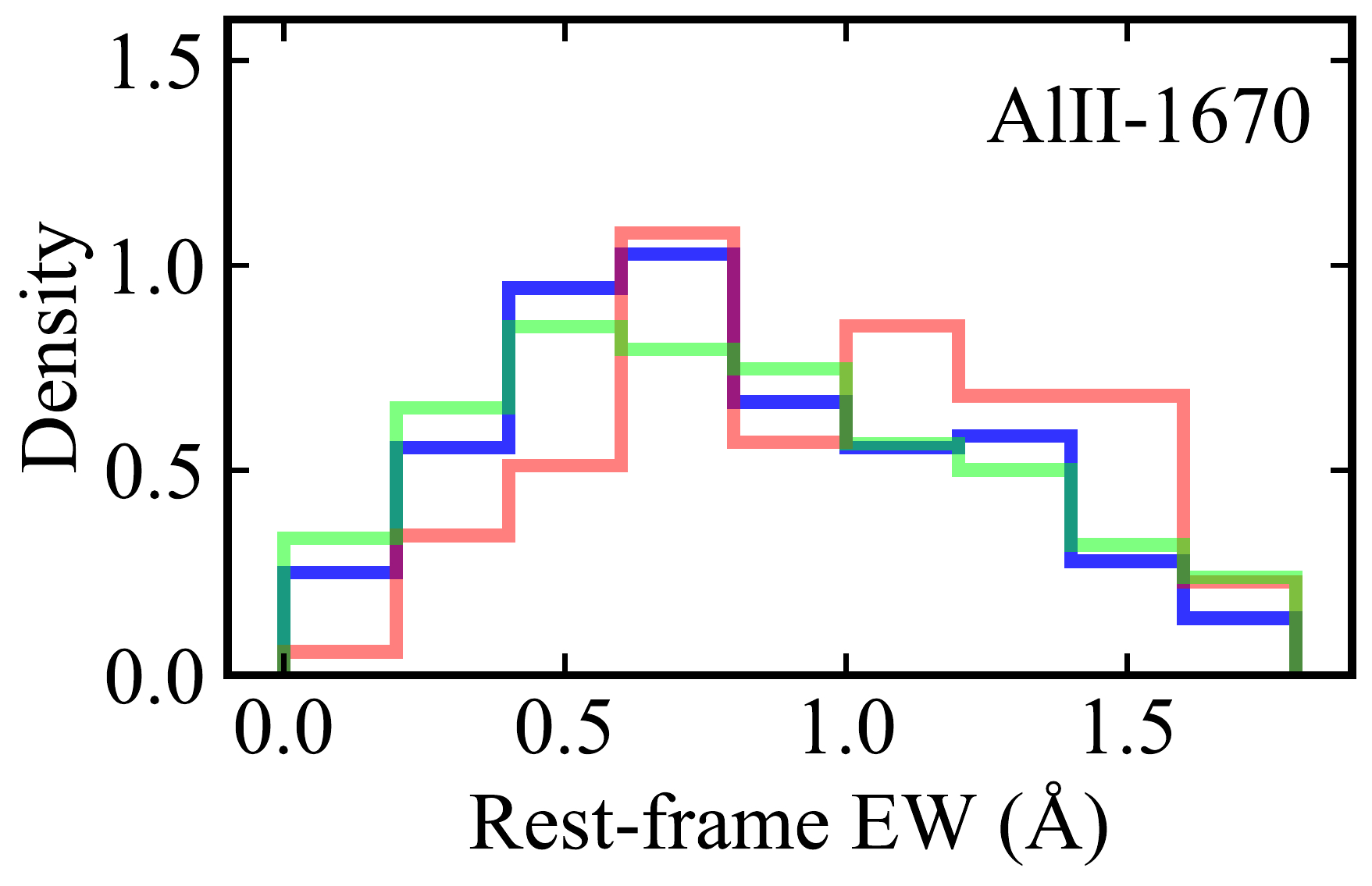}
\includegraphics[width=0.23\hsize]{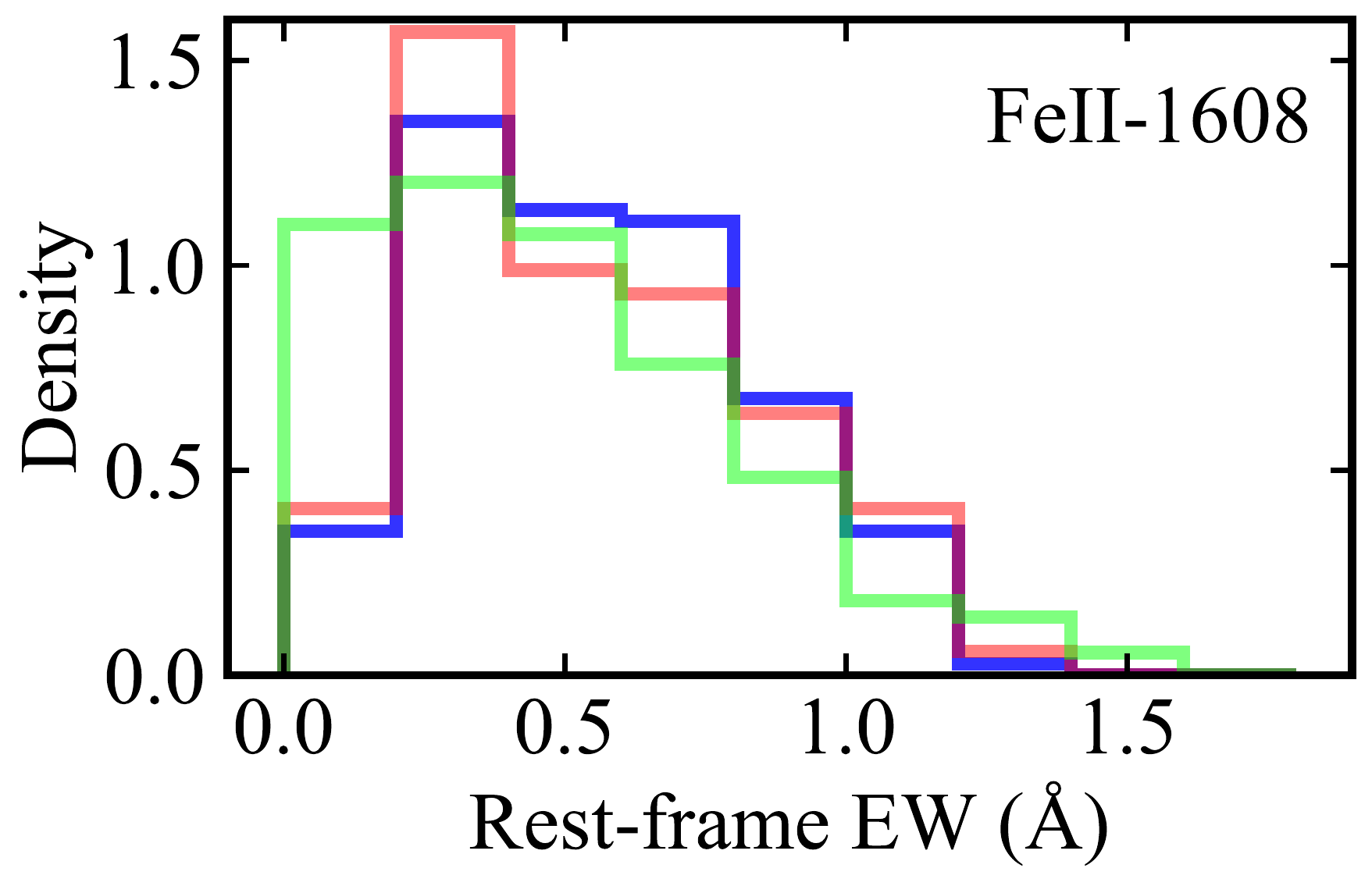} \\
\includegraphics[width=0.23\hsize]{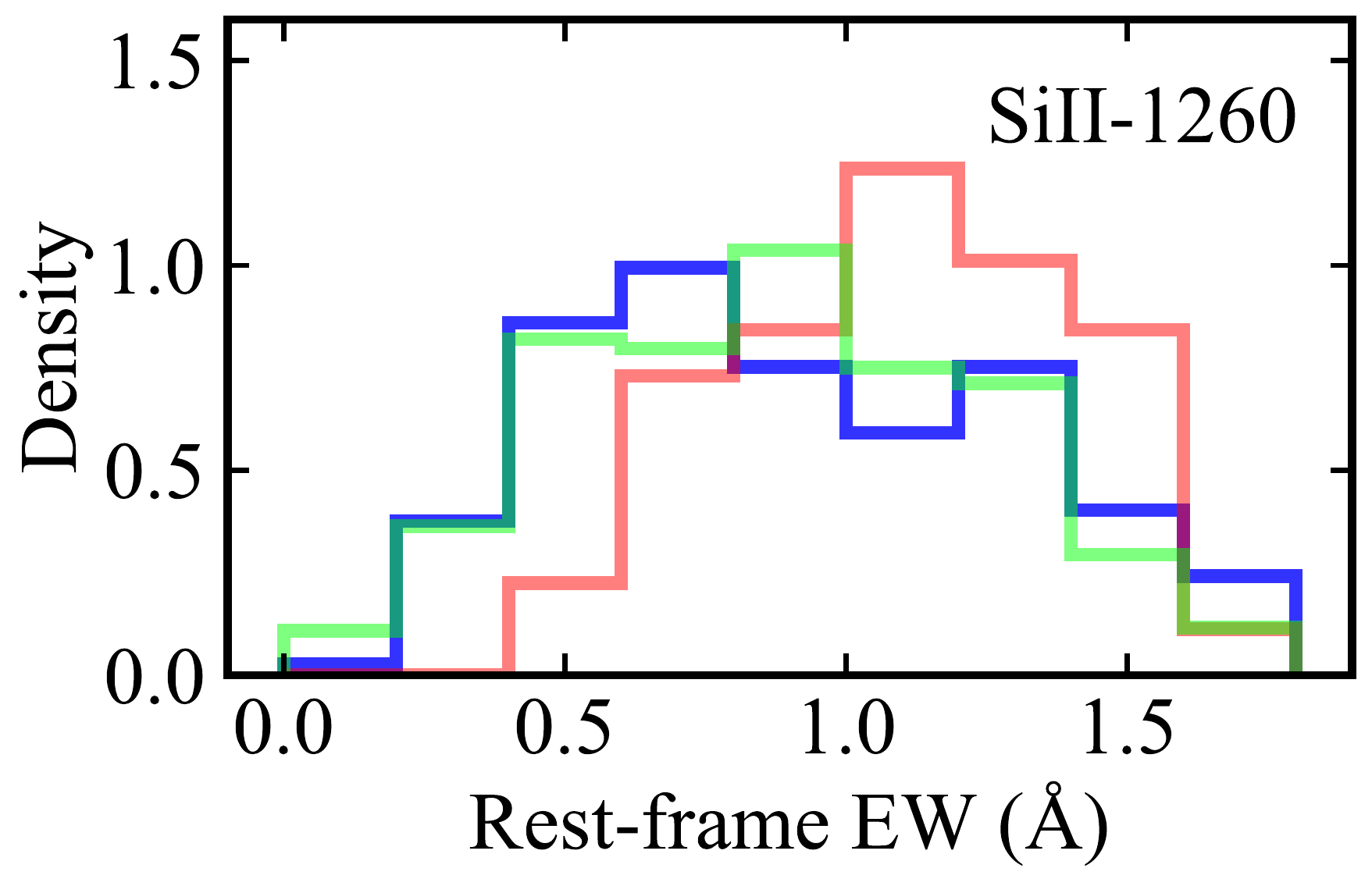}
\includegraphics[width=0.23\hsize]{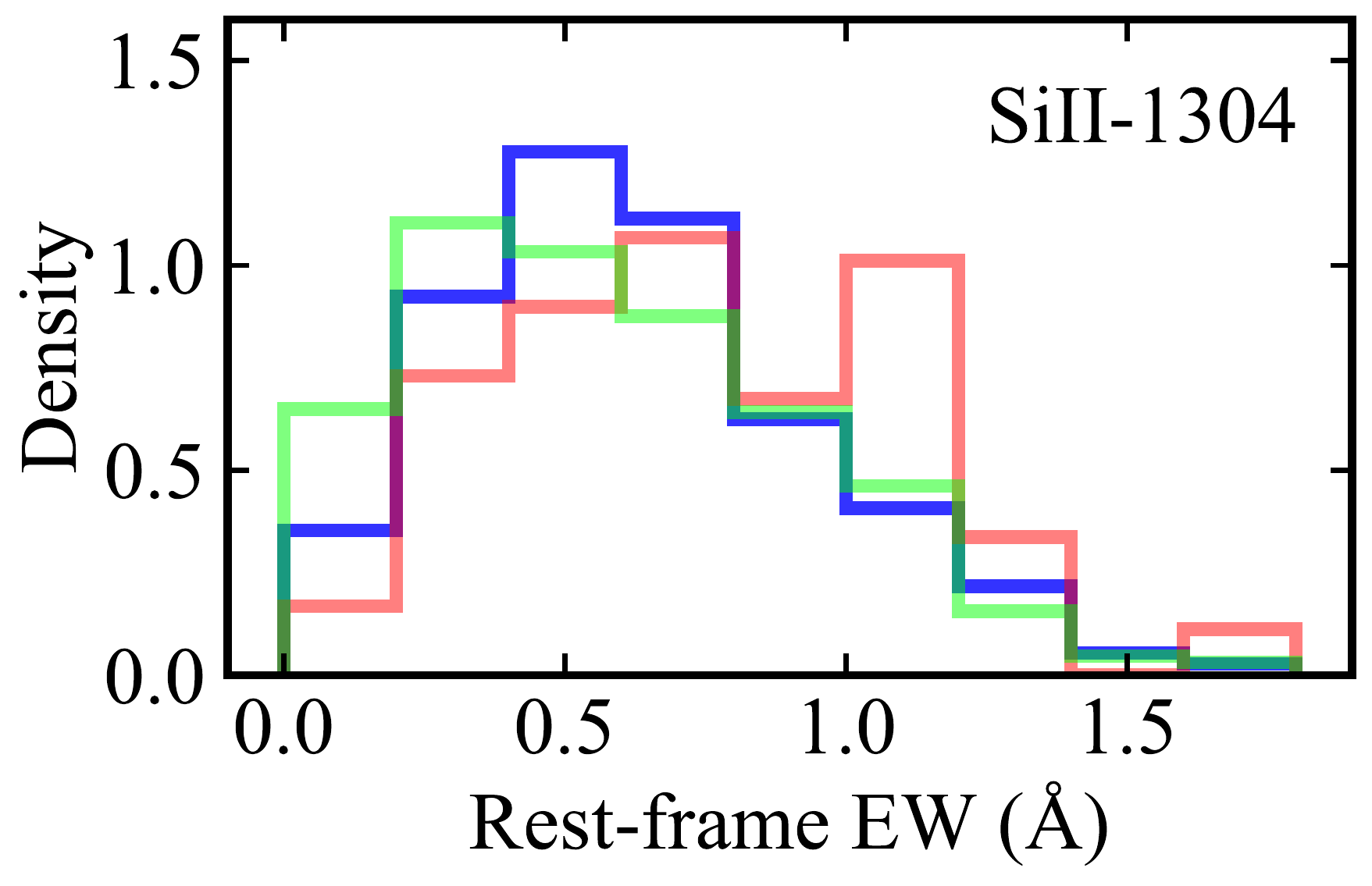}
\includegraphics[width=0.23\hsize]{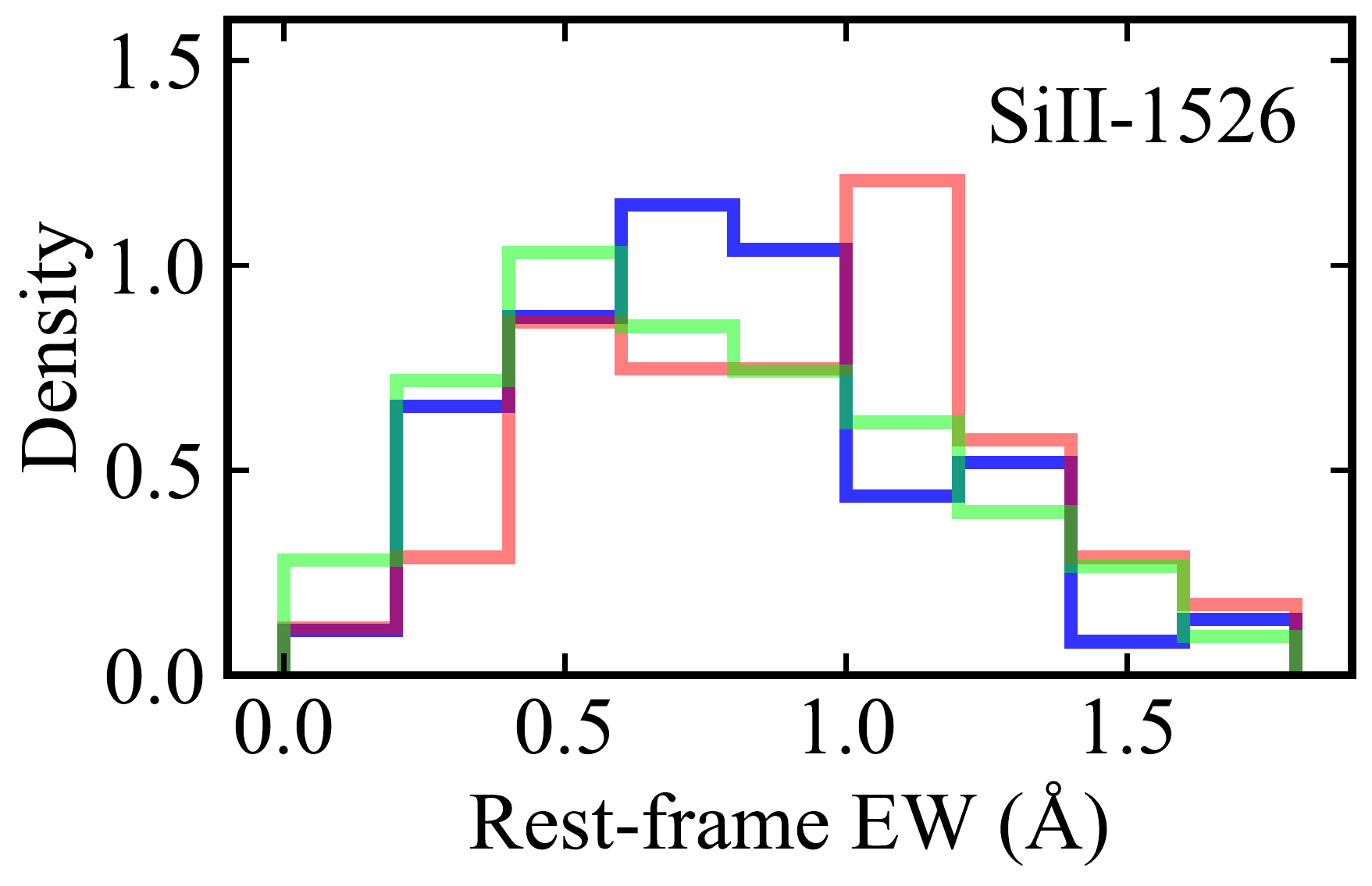}
\includegraphics[width=0.23\hsize]{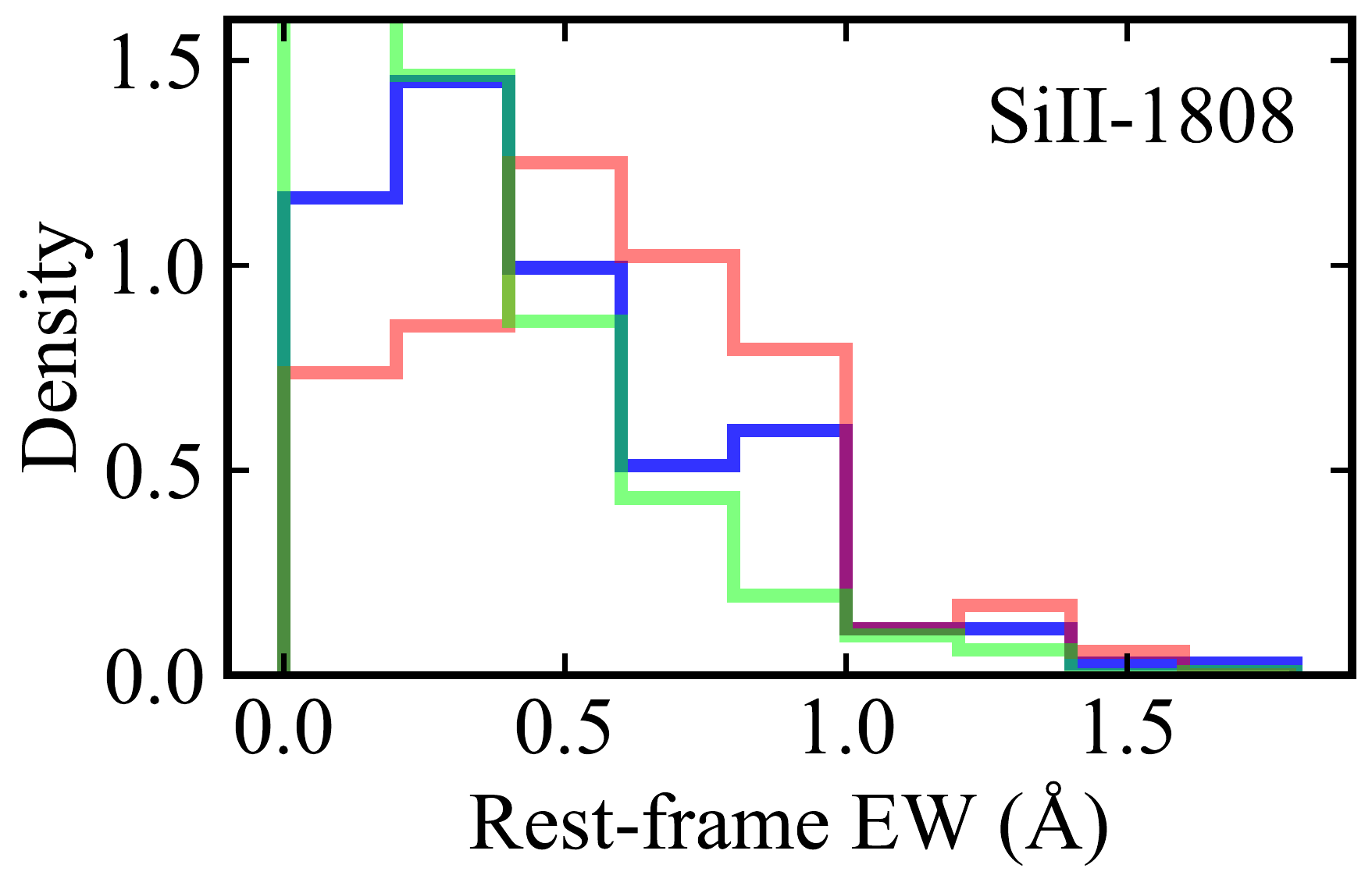} \\
\end{tabular}
\caption{Normalized rest EW distributions of some important species. The blue, green, and red histograms are for DLAs from  \citet{2018MNRAS.476.1151P}, eclipsing DLAs, and ghostly DLAs, respectively.}
 \label{ew_hist}
\end{figure*}

\section{Sample}

\subsection{Training and Test Samples} \label{sec:traintest}

In order to train and test our machine learning models, we need to compile two kinds of quasar spectra: 1) quasar spectra with a DLA located within $\pm$\,1500\,km\,s$^{-1}$ of the quasar redshift (i.e. class\,1 systems), 2) quasar spectra with no absorber at the quasar redshift (i.e. class\,0 systems). Our goal is to find a machine learning model that can distinguish between these two classes. Ghostly DLAs are then found by inspecting the class\,1 systems returned by the machine learning algorithm.

As class\,1 systems, we use the spectra of the 650 quasars compiled by \citet{2018MNRAS.477.5625F}. Each of these quasars has an eclipsing DLA along their line of sight, and the presence of the DLA is visually confirmed. We use the spectra of eclipsing DLAs to train and test our model because it is believed that ghostly DLAs are from the same population as eclipsing DLAs \citep{2017MNRAS.466L..58F,2020ApJ...888...85F}.  If we increase the redshift of each class\,1 eclipsing DLA by 2000\,km\,s$^{-1}$, there will be no absorber at this new redshift in the quasar spectrum. Therefore, we can use the quasar spectra of class\,1 systems, but with the new redshift, as the spectra of our class\,0 systems. This way of defining class\,0 systems has the advantage that both class\,0 and class\,1 systems will have the same distribution of redshift and signal-to-noise ratio (SNR). The absence of any absorber at the redshift of class\,0 systems is also visually confirmed by the lack of absorption from C\,{\sc ii}, C\,{\sc iv}, Si\,{\sc ii}, and Si\,{\sc iv} species. We show in Fig.\,\ref{class_0_1} the stacked spectra of the class\,1 (the black spectrum) and class\,0 (the red spectrum) systems. As expected, no absorption lines are seen at the position of the absorption lines in the class\,0 stacked spectrum.

In total, we have 1300 systems (class\,0 + class\,1 systems) to train the models and evaluate their performances. The k-fold cross validation (k = 5 for this work) is used to evaluate the performance of each model during the training process. Here, the training sample is split into k equal-sized subsamples and each time the model is evaluated on one subsample after it is trained on the remaining k$-$1 samples. This method reduces the error due to the sample variation, and allows model selection and hyper-parameter tuning.

\subsection{Feature Selection} \label{sect:Feature_Selection}

In machine learning, the selection of \emph{features} or input parameters are very important in giving the algorithm the power to discriminate between different classes. In this work, we select the equivalent widths (EWs) and significance levels (SLs) of the absorption from Si\,{\sc ii}\,$\lambda\lambda\lambda\lambda$1260,1304,1526,1808, O\,{\sc i}\,$\lambda$1302, C\,{\sc ii}\,$\lambda$1334, the C\,{\sc iv} doublet, Fe\,{\sc ii}\,$\lambda$1608, the Si\,{\sc iv} doublet, Al\,{\sc ii}\,$\lambda$1670, and the Al\,{\sc iii} doublet transitions as the input parameters. These absorption lines are commonly observed in DLAs, and their EWs can be easily measured from quasar spectra out to redshift of $z_{\rm QSO}$\,$\sim$\,4.3. We also include the median values of the EWs (EW\_med) and SLs (SL\_med) of these absorption lines as two extra features. This increases the total number of input features to 30.

In the case of class\,0 systems, one would expect the EWs and SLs of all absorption lines to be very small and consistent with the noise level. However, it is still possible that some absorption lines from other intervening or associated systems accidentally occur at the wavelengths corresponding to these absorption lines. If this happens, some absorption lines will have high values of EW and SL despite the fact that no absorber is present. Although this issue is inevitable, the EW$\_$med and SL$\_$med are not affected by this. Therefore, we expect these two features to have very high predictive power, as shown later in the paper.

\begin{figure*}
\centering
\begin{tabular}{c}
\includegraphics[width=0.99\hsize]{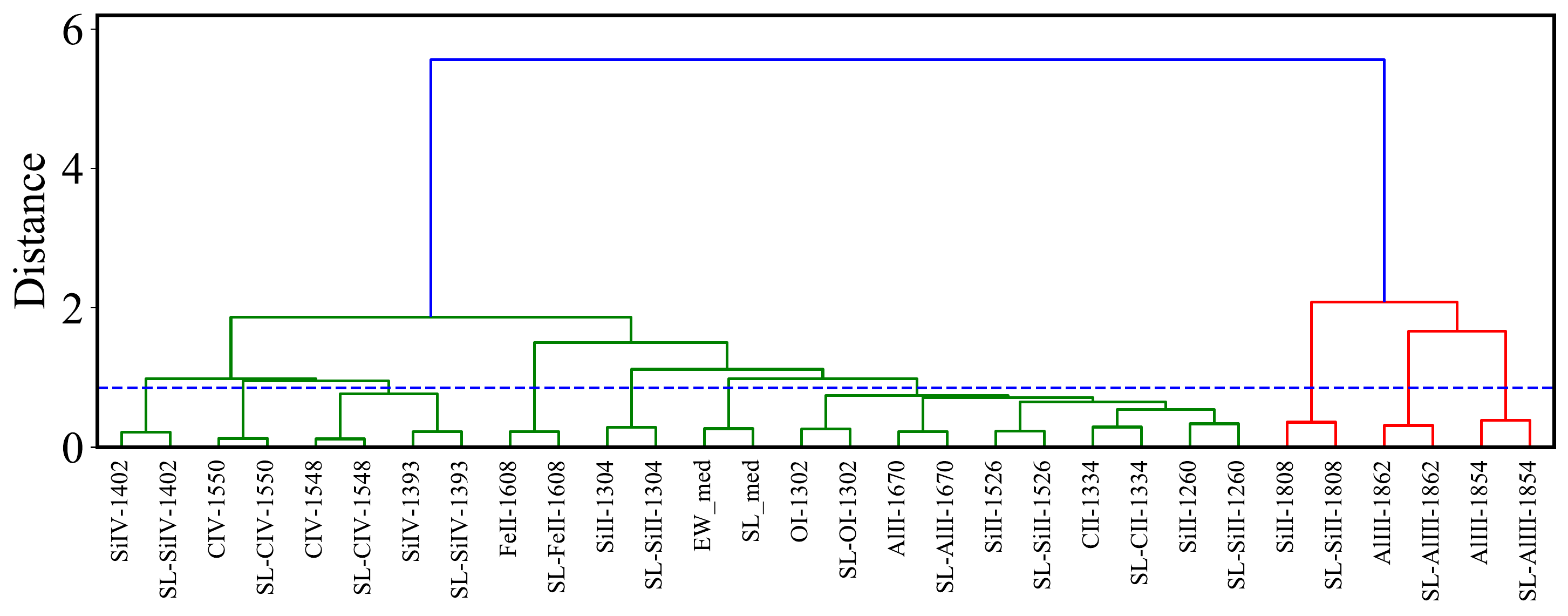}
\end{tabular}
\caption{A dendrogram showing the hierarchical clustering of the input features. The features are clustered based on their respective Spearman correlation strength. The blue horizontal dashed line marks the distance threshold of 0.85. The dendrogram shows that the strongest correlation is found between the EW of each species and its corresponding significance level.}
 \label{dendrogram}
\end{figure*}

\section{Machine Learning Application}

\subsection{Classification Metrics}

After a machine learning classification model is trained, we need to score its performance and estimate its generalization accuracy on unseen data. There are several different metrics to evaluate the performance of a classification model. Here, we only use \emph{accuracy}, \emph{recall}, and \emph{precision}. In binary classification problems such as the current work, \emph{accuracy} is defined as the ratio of the correct predictions on a test data set, i.e.
\begin{equation} \label{eq:1}
accuracy= {\rm \frac{TP + TN}{FN+FP+TP+TN}},
\end{equation}

\noindent
where TP, TN, FN, and FP are \emph{True Positive}, \emph{True Negative}, \emph{False Negative}, and \emph{False Positive}, respectively. True positive (resp. true negative) shows the number of observations that are correctly predicted by the model as class\,1 (resp. class\,0) systems. Moreover, false negative (resp. false positive) denotes the number of observations that belong to class\,1 (resp. class\,0) but are predicted as class\,0 (resp. class\,1) systems. \emph{Recall}, which is given by 

\begin{equation} \label{eq:2}
recall= {\rm \frac{TP }{TP+FN}},
\end{equation}

\noindent
is referred to the percentage of class\,1 systems correctly predicted to belong to class\,1. Finally, \emph{precision}, which is defined as

\begin{equation} \label{eq:3}
precision= {\rm \frac{TP }{TP+FP}},
\end{equation}

\noindent
tells us what proportion of the systems predicted as class\,1 systems actually belong to class\,1.

Predictions of a binary classification model can also be summarized in a confusion matrix with the format shown in Table\,\ref{tableCM1}.
The confusion matrix is actually a visual representation of accuracy, recall, and precision, and it summarizes the number of correct and incorrect predictions for each class.

If we aim at a model with the highest precision, the model will predict lower number of FPs, which means fewer class\,0 systems will be mistakenly predicted as class\,1 systems. As a result, we will have fewer systems for the visual inspection, which is in principle quite good. However, models with higher precision are also more prone to mistakenly predict class\,1 systems as class\,0 systems. This means that high precision models may miss some interesting class\,1 systems. In order to avoid this, we will construct a model that has the highest recall value. This ensures that, compared to models with high precision, fewer class\,1 systems would be missed.

\begin{table}
\caption{Confusion matrix of a 2-class classification problem. The rows correspond to the observed labels and the columns correspond to the labels predicted by the machine learning algorithm.}
\centering 
\setlength{\tabcolsep}{6.8pt}
\renewcommand{\arraystretch}{1.05}
\begin{tabular}{cc|cc}
\multicolumn{2}{c}{}
            &   \multicolumn{2}{c}{Predicted} \\
    &       &   Class 0 &   Class 1              \\ 
    \cline{2-4}
\multirow{2}{*}{\rotatebox[origin=c]{90}{Actual}}
    & Class 0   & TN   & FP                 \\
    & Class 1    & FN    & TP                \\ 
    \cline{2-4}
\end{tabular}
\label{tableCM1}
\end{table}

\begin{figure}
\centering
\begin{tabular}{c}
\includegraphics[width=0.99\hsize]{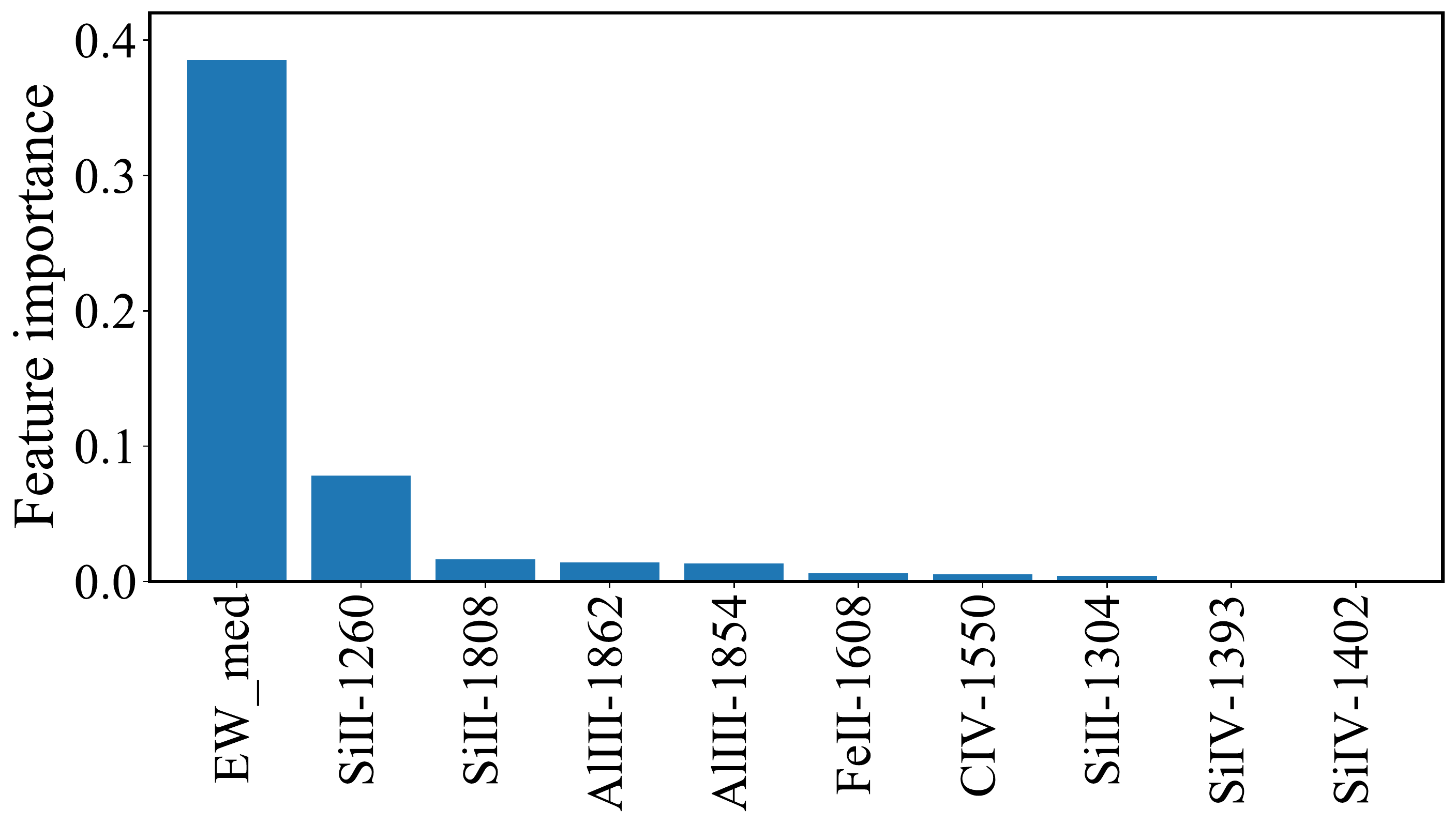}
\end{tabular}
\caption{Importances of the ten input features. The median of the EWs of the absorption lines (i.e. EW\_med) has the highest importance.}
 \label{permutation}
\end{figure}

\subsection{Model Training and Performance}

In this work, we use the XGBoost algorithm implemented in scikit-learn v0.19.1 \citep{JMLR:v12:pedregosa11a} in order to build a model capable of classifying quasar spectra into class\,0 and class\,1 categories (see section\,\ref{sec:traintest}). XGBoost is a decision-tree-based ensemble machine learning algorithm that is based on a gradient boosting framework. The algorithm builds strong models by constructing and merging a multitude of decision trees.

An important issue in any machine learning application is to avoid overfitting and under-fitting. Overfitting occurs when a model learns patterns that are too specific to the training data, resulting in weak generalization of the model to other data sets. On the other hand, under-fitting occurs when the model is too simple  to adequately capture the underlying structure of the data. To avoid these issues, one needs to properly fine-tune the hyperparameters of the model. For example, in XGBoost, \emph{max\_depth}, which sets the maximum depth of the decision tree, is one of the several hyperparameters that control the overfitting of the model. Setting a high value for \emph{max\_depth} may increase the likelihood of overfitting as deeper decision trees are prone to capture patterns that are too specific to the training data. 

We, therefore, perform a grid-search on the key hyperparameters to obtain the optimal XGBoost model. The grid-search uses five-fold cross-validation to train and test the models, and the recall score is adopted as the performance metric. Although XGBoost has many hyperparameters to adjust, we only tune those which are more important. The hyperparameters and their optimal values are listed in Table\,\ref{table1}. The accuracy, recall, and precision of our optimal model are 0.987\,$\pm$\,0.007, 0.986\,$\pm$\,0.003, and 0.988\,$\pm$\,0.014, respectively, and the corresponding confusion matrix is shown in Table\,\ref{tableCM2}.

\begin{table}
\caption{Confusion matrix constructed using our best XGBoost model.}
\centering 
\setlength{\tabcolsep}{6.8pt}
\renewcommand{\arraystretch}{1.05}
\begin{tabular}{cc|cc}
\multicolumn{2}{c}{}
            &   \multicolumn{2}{c}{Predicted} \\
    &       &   Class 0 &   Class 1              \\ 
    \cline{2-4}
\multirow{2}{*}{\rotatebox[origin=c]{90}{Actual}}
    & Class 0   & 642   & 8                 \\
    & Class 1    & 9    & 641                \\ 
    \cline{2-4}
\end{tabular}
\label{tableCM2}
\end{table}

From the confusion matrix, we can see that out of 641\,+\,9 class\,1 systems, the model correctly identifies 641 as class\,1 systems and 9 erroneously as class\,0 systems. Similarly, out of 642\,+\,8 class\,0 systems, the model correctly identifies 642 as class\,0 systems and 8 erroneously as class\,1 systems. We visually inspected all 9 false negative class\,1 systems and found that all of them have very weak metal absorption lines. Figure\,\ref{FNStack} shows the median-stacked spectrum of these 9 false negative systems as the black spectrum, the median-stacked spectrum of the 641 true positive systems as the red spectrum, and the 95th percentile stacked spectrum of the true positive systems as the blue spectrum.

The blue spectrum in Fig.\,\ref{FNStack} is constructed in the same way as the red spectrum except that instead of using the median which is the 50th percentile, we use the 95th percentile of the 641 true positive class\,1 systems. Since the 95th percentile is used here, absorption systems with the weakest absorption lines have the dominant contribution in this stacked spectrum. Interestingly, absorption lines from the false negative stacked spectrum are still much weaker than those in the 95th percentile stacked spectrum. The strength of the absorption lines in the 95th percentile stacked spectrum shows that the majority of class\,1 systems have rather strong metal absorption lines. Therefore, our machine learning model, which is trained on these rather strong metal absorbers, may not be very sensitive in detecting systems with very weak metal absorption lines. It would therefore be interesting to look for weak ghostly DLA absorbers by training a machine learning model on a sample of DLAs with very weak metal absorption lines.

\begin{figure}
\centering
\begin{tabular}{c}
\includegraphics[width=0.99\hsize]{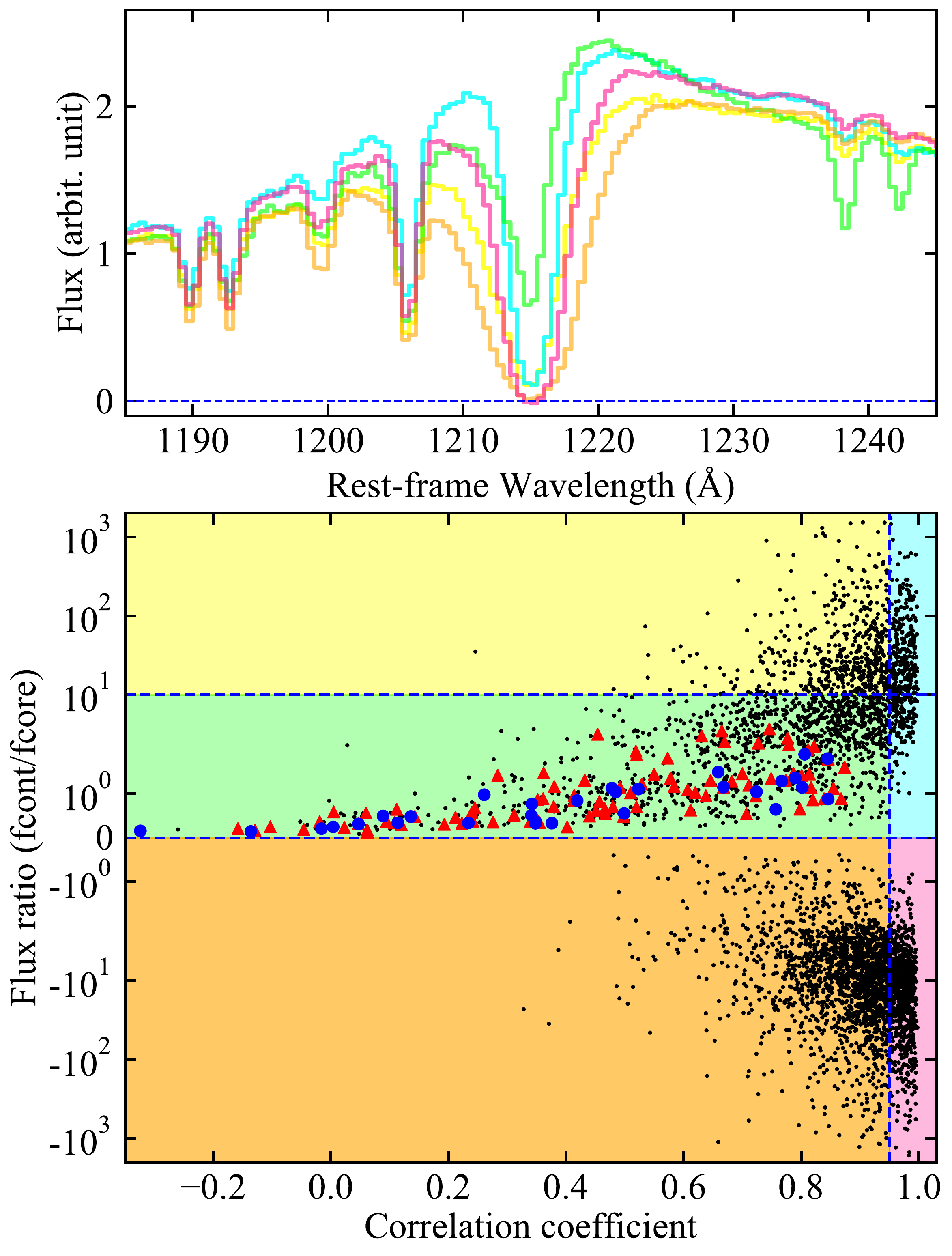}
\end{tabular}
\caption{\emph{Lower panel:} The continuum-to-core flux ratio as a function of the correlation coefficient for all the systems from the $S_{1}$ sample. The blue dots show the data points of the 30 ghostly absorbers from \citetalias{2020ApJ...888...85F}, and the red triangles show the newly found ghostly DLAs. The two horizontal lines mark the continuum-to-core flux ratio of 0 and 10, and the vertical line shows the correlation coefficient of 0.95. \emph{Upper panel:} the median stacked spectra of the systems residing in each colored region from the lower panel. Each spectrum has the same color as its corresponding region.}
 \label{fratiovscoeff}
\end{figure}

\subsection{Testing the model using non-eclipsing DLAs}

In this section, we further test the performance of our model using DLAs with velocity separation $\Delta$$V$\,$>$\,2000\,km\,s$^{-1}$ with respect to their background quasars. Due to the higher velocity separation, these DLAs do not eclipse their background quasars.  There are several DLA catalogs available in the literature that can be used for this purpose \citep{2004PASP..116..622P,2005ApJ...635..123P,2012A&A...547L...1N,2018MNRAS.476.1151P,2020MNRAS.496.5436H}. Here, we use the DLA sample from \citet{2018MNRAS.476.1151P}. These DLAs are found through the detection of the strong Ly$\alpha$ absorption in the spectra. In contrast, we search for DLAs using a metal absorption line template. Since the former approach is less sensitive to the quality of the spectra, it generally finds more DLAs compared to the latter approach \citep[see section\,2.2 in][]{2018MNRAS.477.5625F}. However, the two approaches mostly converge for DLAs with log\,$N$(H\,{\sc i})\,$>$\,21.0 \citep[see figure\,3 in][]{2018MNRAS.477.5625F}.

Therefore, to test our model, we take into account only those DLAs with log\,$N$(H\,{\sc i})\,$>$\,21.0. We also exclude DLAs with $\Delta$$V$\,$>$\,10\,000\,km\,s$^{-1}$ to ensure that most of the absorption lines of interest lie outside the Ly$\alpha$ forest, SNR\,$<$\,4, and non-zero balnicity index \citep{1991ApJ...373...23W} to reject broad absorption line (BAL) quasars. Applying these constrains on the DLA catalog of \citet{2018MNRAS.476.1151P} returns 186 DLAs. These DLAs are labeled as class\,1 systems. We then shift the redshift of these DLAs by 2000\,km\,s$^{-1}$ to create the class\,0 systems (see section\,\ref{sec:traintest}). Finally, we apply our XGBoost model on these 372 systems. The resulting confusion matrix is shown in Table\,\ref{parksCM}. The high accuracy of 97.3\,\% further shows that our model should perform well when applied on SDSS-DR14 spectra.

We show in Fig.\,\ref{ew_hist} the EW distributions of different species detected in eclipsing DLAs (green histograms), the DLAs from \citet{2018MNRAS.476.1151P} that are used in this section (blue histograms), and ghostly DLAs (red histograms). As shown in this figure, there is no striking difference in the EW distribution of metal lines in the eclipsing DLAs and the DLAs from \citet{2018MNRAS.476.1151P}. This implies that machine learning models trained on eclipsing DLAs could also be used to find non-eclipsing DLAs.

As shown in Fig.\,\ref{ew_hist}, the low ionization species in ghostly DLAs exhibit very similar EW distributions compared to the other two DLAs. The higher C\,{\sc ii} EWs in ghostly DLAs is mainly due to the fact that it is blended with C\,{\sc ii}$^{*}$\,$\lambda$1335 which is generally stronger in ghostly DLAs \citepalias{2020ApJ...888...85F}. On the other hand, high-ionization species (Al\,{\sc iii}, Si\,{\sc iv}, and C\,{\sc iv}) are also stronger in ghostly DLAs (see upper panels in Fig.\,\ref{ew_hist}). Stronger high ionization species in ghostly DLAs implies higher level of ionization of their outer layers due to the proximity to the quasar \citepalias{2020ApJ...888...85F}.

Motivated by the apparent differences seen in the EWs of metal absorption lines in ghostly and other DLAs, we tried to construct a machine learning model that could exploit these differences and \emph{directly} distinguish between these DLAs. Unfortunately, the resulting model has a low \emph{recall} value ($\sim$\,60\,\%), and hence it does not perform well in distinguishing between ghostly and other DLAs. The poor performance of the model implies that one cannot distinguish between these DLAs solely based on their metal absorption line properties. 

\begin{table}
\caption{Confusion matrix constructed using our best XGBoost model on DLAs from \citet{2018MNRAS.476.1151P}.}
\centering 
\setlength{\tabcolsep}{6.8pt}
\renewcommand{\arraystretch}{1.05}
\begin{tabular}{cc|cc}
\multicolumn{2}{c}{}
            &   \multicolumn{2}{c}{Predicted} \\
    &       &   Class 0 &   Class 1              \\ 
    \cline{2-4}
\multirow{2}{*}{\rotatebox[origin=c]{90}{Actual}}
    & Class 0   & 184   & 2                 \\
    & Class 1    & 8    & 178                \\ 
    \cline{2-4}
\end{tabular}
\label{parksCM}
\end{table}

\begin{figure}
\centering
\begin{tabular}{c}
\includegraphics[width=0.99\hsize]{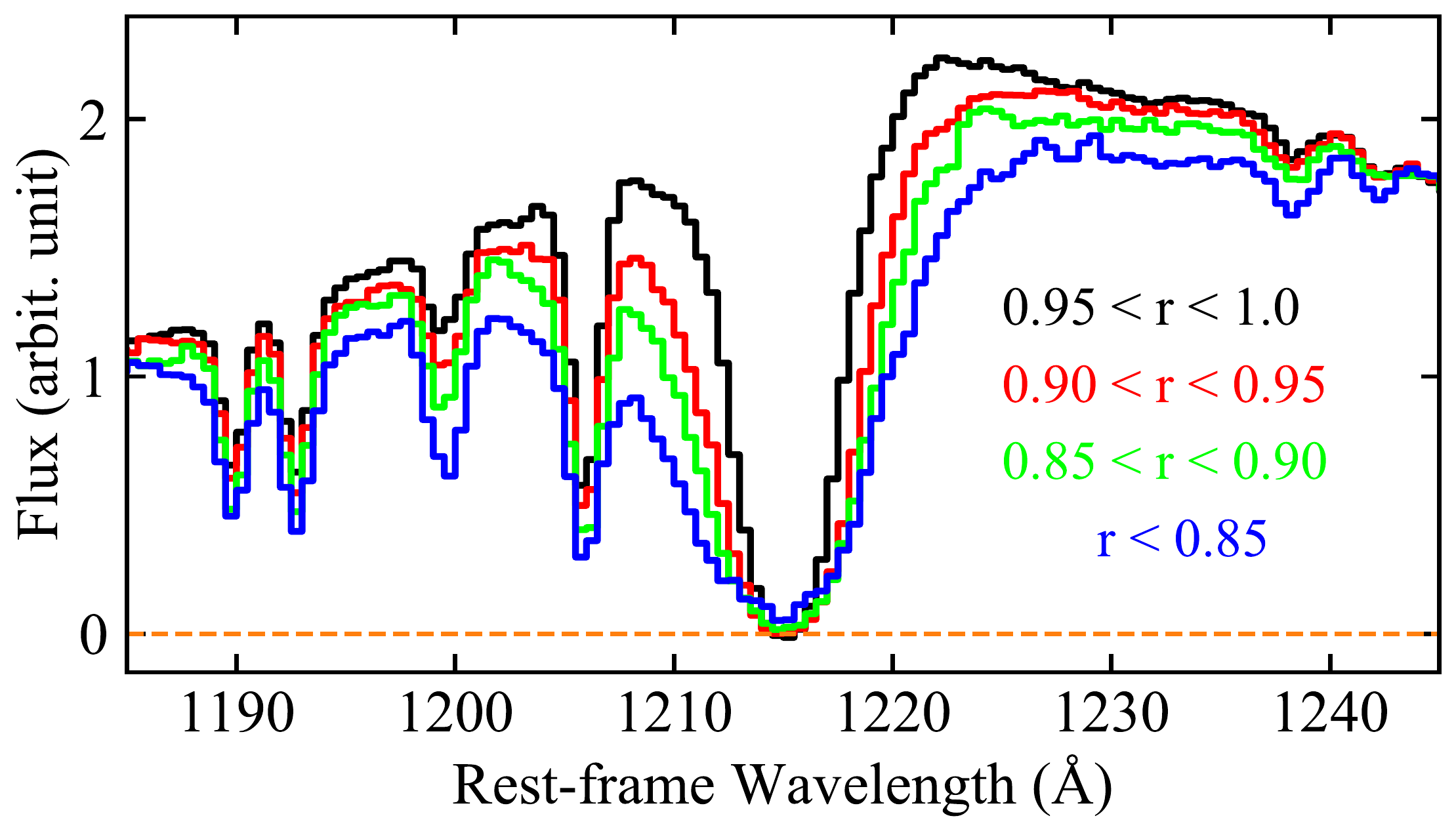}
\end{tabular}
\caption{The stacked spectra of the systems with different correlation coefficients. To construct these stacked spectra, only systems with the continuum-to-core flux ratio $\le$\,0 are taken into account.}
 \label{NHIvscoeff}
\end{figure}

\subsection{Importance of Individual Input Features}

We can measure the distinguishing power of each of our 30 input features by looking at its feature importance. Tree-based machine learning algorithms like XGBoost provide feature importances out of the box. However, here we use the \emph{permutation} technique, which is a more reliable approach in measuring the importance of a feature. The permutation importance of a feature is defined as the decrease in the model accuracy when the values of that feature are randomly shuffled. Shuffling the feature values removes any relationship that might have existed between the feature and the target. Therefore, any drop in the model accuracy could be an indicator of how much the model is dependent on the feature.

When two features are correlated, the permutation of one feature will have little effect on the performance of the model as it could still have access to the feature through its correlated feature. As a result, lower importance will be assigned to both features although they might actually be very important. To overcome this problem, we use hierarchical clustering to cluster features that are strongly correlated \citep{JMLR:v12:pedregosa11a}. The result of the hierarchical clustering of our 30 features is shown in Fig.\,\ref{dendrogram} in the form of a \emph{dendrogram}. A dendrogram is a diagram that shows the hierarchical correlation between different features. Here, the Spearman rank-order correlation is employed to measure the correlations, and the \emph{ward} linkage is adopted to measure the distance.

As shown in Fig.\,\ref{dendrogram}, the strongest correlation is found between the EW of each species and its corresponding significance level. In the left-hand side of the dendrogram, we can see that high ionization species are clustered together. This is indeed very impressive as the clustering algorithm was not provided any physics principles but rather discriminate between high and low ionization species. Moreover, Si\,{\sc ii}\,$\lambda$1808 and the Al\,{\sc iii} doublet cluster with neither low nor high ionization species, which means they are not strongly correlated with those transitions. Their lower correlation is due to the fact that they are intrinsically weak absorption lines.

By picking a distance threshold of 0.85 in the dendrogram, we end up with 10 separate clusters (see Fig.\,\ref{dendrogram}). The choice of the threshold is rather arbitrary, and mostly depends on the domain knowledge we have about the data. Here, we choose a threshold of 0.85 for two reasons: 1) at this threshold, SL$\_$med and EW$\_$med form a cluster of their own, allowing us to evaluate their importances against other features, and 2) most of the strong low ionization species cluster together at this threshold. Since features inside a cluster are strongly correlated, we keep only a single feature from each cluster. As a result, we are left with only 10 features. The permutation importances of these 10 features are shown in Fig.\,\ref{permutation}.

\begin{figure}
\centering
\begin{tabular}{c}
\includegraphics[width=0.99\hsize]{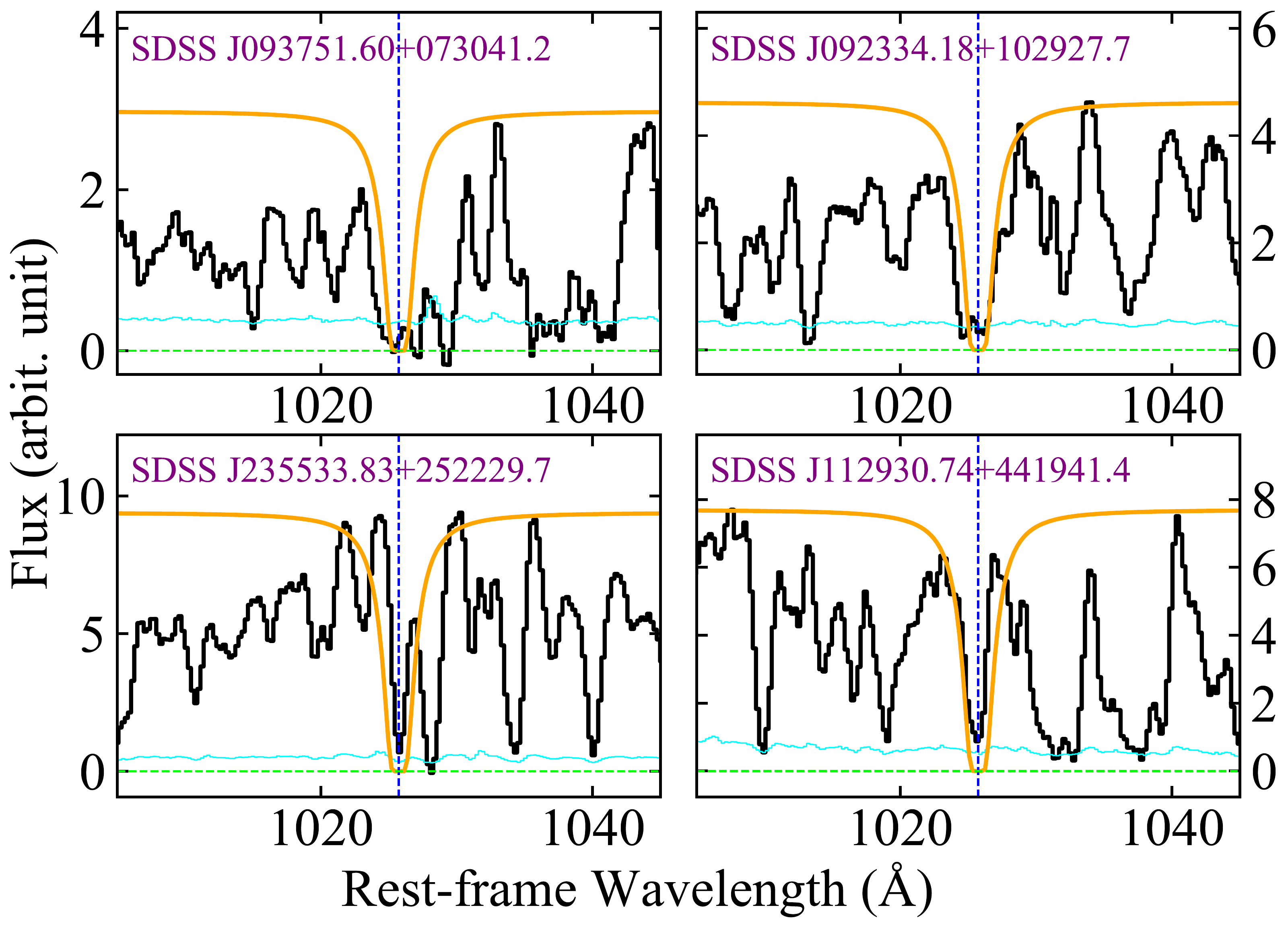}
\end{tabular}
\caption{The Ly$\beta$ spectral regions of four ghostly absorbers. In each panel, the red curve shows a synthetic Ly$\beta$ absorption line with log\,$N$(H\,{\sc i})\,=\,21.0 and $b$-value of 100\,km\,s$^{-1}$. The upper panels (resp. lower panels) show Ly$\beta$ absorption that are broader (resp. narrower) than the synthetic spectrum.}
 \label{goodbad}
\end{figure}

\begin{figure*}
\centering
\begin{tabular}{c}
\includegraphics[width=0.99\hsize]{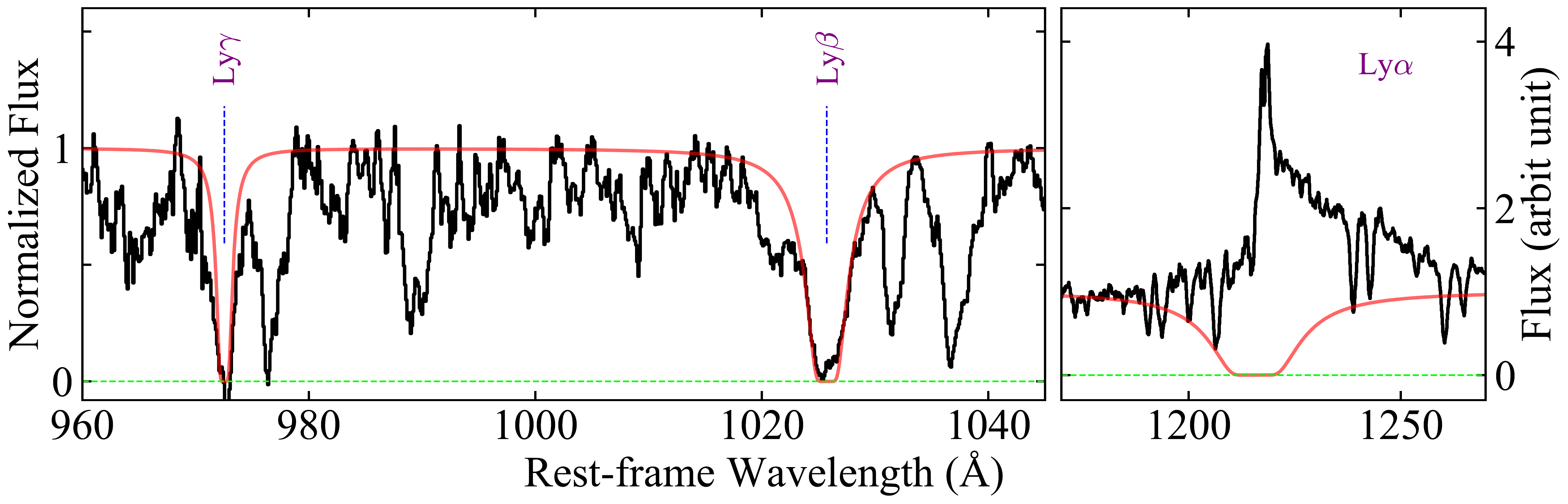}
\end{tabular}
\caption{Voigt profile fit (red curves), with log\,$N$(H\,{\sc i})\,=\,21.5 and $b$-value of 50\,km\,s$^{-1}$, on the Ly$\beta$ and Ly$\gamma$ absorption lines in the stacked spectrum of the ghostly absorbers with strong Ly$\beta$ absorption lines (see the text). The H\,{\sc i} column density is constrained by the strong damping wings of the Ly$\beta$ absorption line.}
 \label{lyabg}
\end{figure*}

\section{Results}

In this section, we apply our best XGBoost model to the SDSS DR14 spectra. Here, we take into account only quasars with $z_{\rm QSO}$\,$>$\,2.0 to make sure that the Ly$\alpha$ spectral region falls in the observed spectral window, balnicity index of zero to reject BAL quasars, ZWARNING\,=\,0 \citep{2018A&A...613A..51P} to exclude quasars with highly uncertain redshifts, and SNR\,$>$\,4.0. The SNR is measured over the spectral region between 1410 to 1510\,\textup{\AA} in the quasar rest frame. Applying these constraints on the SDSS DR14 spectra leaves us with 155\,851 quasars. We call this sample of quasars the $S_{0}$ sample. We will search for candidate metal absorbers in the $S_{0}$ sample, measure the EWs and SLs of their absorption lines, and then apply our best XGBoost model on these measurements.

As shown in Fig.\,\ref{permutation}, EW$\_$med has the highest importance. The general trend in Fig.\,\ref{permutation} is that low ionization species have higher importance than high ionization species. Si\,{\sc ii}\,$\lambda$1304 has a slightly lower importance than C\,{\sc iv}\,$\lambda$1550, and this is because most of its importance is shared with Si\,{\sc ii}\,$\lambda$1260 which is highly correlated with it. The high ionization species also have very low importance, indicating that similar to Si\,{\sc ii}\,$\lambda$1304, most of their importance has been shared with low ionization species. Nevertheless, we use all 30 features when training and evaluating our machine learning model.

\subsection{Search for Metal Absorbers}

Since ghostly (DLA) absorbers reveal almost no signature of an Ly$\alpha$ absorption in the spectrum, one needs to rely on the presence of metal absorption lines in the spectrum in order to identify them. We therefore use a Spearman correlation analysis to search for candidate metal absorption line systems. For this purpose, we create a metal absorption line template using 12 Gaussian functions. The Gaussian functions represent absorption from the metal line transitions listed in section\,\ref{sect:Feature_Selection}. We note that the C\,{\sc iv} doublet transitions are not included because the absorption from the two lines of the doublet are usually severely blended with each other.

For each quasar in the $S_{0}$ sample, its spectrum is cross-correlated against the metal absorption line template, and the redshift at which the correlation function reaches its maximum is taken as the absorber redshift. Then, using this absorption redshift, we calculate the EWs and SLs of the absorption lines listed in section\,\ref{sect:Feature_Selection}. Similar to \citetalias{2020ApJ...888...85F}, the search for metal absorbers is conducted within 1500\,km\,s$^{-1}$ of the quasar emission redshift. The measured EWs and SLs will be used to distinguish between class\,0 and class\,1 systems.

\subsection{Applying the Model to SDSS-DR14 Spectra}

By applying our best XGBoost model on the $S_{0}$ sample, 4804 systems are identified as class\,1 systems. We call this sample, the $S_{1}$ sample. In principle, the $S_{1}$ sample contains Lyman Limit Systems (LLSs, absorbers with log\,$N$(H\,{\sc i})\,$\le$\,19), sub-DLAs (absorbers with 19\,$\le$\,log\,$N$(H\,{\sc i})\,$\le$\,20.3), DLAs, ghostly absorbers, and some false positive detection. Our goal is to separate ghostly absorbers from the other systems. Below, we explain how this is done.

Since ghostly absorbers, by definition, exhibit no apparent Ly$\alpha$ absorption in their spectra, correlation of a synthetic Ly$\alpha$ absorption with the Ly$\alpha$ spectral region of ghostly and non-ghostly absorbers should yield different correlation coefficients. These differing correlation coefficients could help us distinguish between ghostly and non-ghostly absorbers. Moreover, the ratio of the quasar flux at $\lambda$\,$\sim$\,1300\,\textup{\AA} (i.e. quasar continuum around Ly$\alpha$ spectral region) to the flux at $\lambda$\,$\sim$\,1215\,\textup{\AA} (i.e. the Ly$\alpha$ absorption core) is also different in ghostly and non-ghostly absorbers. This is because in ghostly absorbers the flux at $\lambda$\,$\sim$\,1215\,\textup{\AA} never reaches zero, due to the leaked Ly$\alpha$ emission from the BLR. 

For each quasar, the minimum flux of the spectral pixels located within 300\,km\,s$^{-1}$ of the absorption redshift is taken as the Ly$\alpha$ absorption core flux. Moreover, the correlation coefficient associated with each quasar in the $S_{1}$ sample is found by cross-correlating the Ly$\alpha$ spectral region of the quasar spectrum with a series of synthetic Ly$\alpha$ absorption profiles with $N$(H\,{\sc i}) in the range 19\,$\le$\,log\,$N$(H\,{\sc i})\,$\le$\,22. 
We then adopt the maximum coefficient as the correlation coefficient of the quasar. Here, we exploit these two distinguishing parameters (i.e. the correlation coefficient and the continuum-to-core flux ratio) to separate candidate ghostly and non-ghostly absorbers.

The lower panel in Fig.\,\ref{fratiovscoeff} shows the continuum-to-core flux ratio as a function of the correlation coefficient for all the systems from the $S_{1}$ sample. Here, the blue dots show the data points of the 30 ghostly absorbers from \citetalias{2020ApJ...888...85F}. The correlation coefficient of the blue dots ranges from $-$0.32 to 0.85, and their continuum-to-core flux ratio from 0.14 to 1.9. Motivated by these numbers, we constrain our search for candidate ghostly absorbers to those systems for which the continuum-to-core flux ratio ranges from 0 to 10, and the correlation coefficient from $-$1.0 to 0.95. Here, we chose a more extended range for each of these parameters compared to those of the blue dots. The parameter space defined by these numbers is shown as the green region in the lower panel of Fig.\,\ref{fratiovscoeff}.

The green region in Fig.\,\ref{fratiovscoeff} contains 1124 systems, and the median SNR of these systems around the Ly$\alpha$ spectral region (i.e. 1320 to 1380\,\textup{\AA}) is $\sim$\,8.9. These systems are our candidate ghostly absorbers. This region contains four kinds of systems: 1) LLSs (at the SDSS spectral resolution, the core of the Ly$\alpha$ absorption in LLSs does not reach zero flux), 2) DLAs with some narrow Ly$\alpha$ emission in their trough (the presence of the narrow Ly$\alpha$ emission increases the flux at the core of the DLA, resulting in a decrease in the continuum-to-core flux ratio), 3) ghostly absorbers, and 4) some false positive detection. By visually inspecting all these systems, we found 89 ghostly absorbers. We call this the $S_{G}$ sample. Out of these 89 ghostly absorbers, 59 of them are new systems. These new ghostly DLAs are shown as red triangles in the lower panel of Fig.\,\ref{fratiovscoeff}. The SDSS name, $z_{QSO}$, and $z_{abs}$ of all 89 ghostly absorbers are listed in Table\,\ref{newghostly}. 41 absorbers in the $S_{G}$ sample also cover the Ly$\beta$ spectral region. We call this the $S_{\beta}$ sample. Before going into any further analysis of these ghostly absorbers, we first comment on the systems residing in the remaining colored regions in Fig\,\ref{fratiovscoeff}. The median-stacked spectrum of the systems in each colored region in Fig.\,\ref{fratiovscoeff} is shown in the upper panel of the figure as a curve with a similar color.

The number of systems in the cyan region in Fig.\,\ref{fratiovscoeff} is 437 with the median SNR of 16.8. As shown in Fig.\,\ref{fratiovscoeff}, the Ly$\alpha$ absorption in the stacked spectrum of the systems located in the cyan region shows a symetric line profile. This  symmetry implies that the Ly$\alpha$ absorption of the majority of the systems in this region are unaffected by intervening absorption and/or noise. This is not surprising as these systems have rather high correlation coefficients ($r$\,$\ge$\,0.95).

The number of systems in the yellow region in Fig.\,\ref{fratiovscoeff} is 522 with the median SNR of 13.4. The Ly$\alpha$ absorption in the stacked spectrum of these systems shows some asymmetry, especially on the blue wing (see Fig.\,\ref{fratiovscoeff}). This asymmetry is an indication that the blue wing of the Ly$\alpha$ absorption in most of the systems are suffering from contaminating absorption. The fact that the asymmetry is more prominent on the blue wing implies that the contaminating features are mostly absorption from the Ly$\alpha$ forest.

The pink region in Fig\,\ref{fratiovscoeff} contains 830 systems with the median SNR of 14.5. The negative continuum-to-core flux ratio in these systems indicates that the core of their Ly$\alpha$ absorption reaches zero flux. As shown in Fig.\,\ref{fratiovscoeff}, the Ly$\alpha$ absorption in the stacked spectrum of these systems is broader than that of the cyan regions. This implies that most absorbers in the pink region have higher H\,{\sc i} column density. 

The number of systems in the orange region in Fig.\,\ref{fratiovscoeff} is 1891 with the median SNR of 9.1. Similar to the systems in the pink region, the negative continuum-to-core flux ratio implies that the core of the Ly$\alpha$ absorption in these systems also reaches zero flux. As shown in Fig.\,\ref{fratiovscoeff}, the Ly$\alpha$ absorption in the stacked spectrum of these systems is broader than that of the pink region. This implies that the fraction of systems with higher H\,{\sc i} column density in the orange region is higher than in the pink region. 

Since the blue wing of an absorber with higher H\,{\sc i} column density could extend more into the Ly$\alpha$ forest region, it has a higher chance of getting contaminated by the Ly$\alpha$ forest absorption lines. Therefore, we expect absorbers with higher H\,{\sc i} column density to have smaller correlation coefficients compared to those with lower H\,{\sc i} column density. To check this, we create stacked spectra using quasar absorbers with different correlation coefficients, and then compare the strength of their Ly$\alpha$ absorption lines. For this exercise, we take into account only the systems located in the orange and pink regions. The result is shown in Fig.\,\ref{NHIvscoeff}. As shown in this figure, absorbers with stronger Ly$\alpha$ absorption have lower correlation coefficients. This explains why the orange Ly$\alpha$ absorption is broader than the pink one in the upper panel of Fig.\,\ref{fratiovscoeff}.

\begin{table}
\caption{Newly found ghostly absorbers. First column: J2000 coordinates of the quasars. Second column: quasar emission redshift. Third column: ghostly absorber absorption redshift. The full version of this table is available on-line. Only the first 10 entries are shown here.}
\centering 
\setlength{\tabcolsep}{9.8pt}
\renewcommand{\arraystretch}{1.05}
\begin{tabular}{c c c} 
\hline\hline
SDSS Name & $z_{QSO}$ & $z_{abs}$  \\
\hline 
000125.69+141022.3   &   2.421   &   2.416  \\
000943.73+132032.6   &   2.390   &   2.376  \\
000958.65+015755.1   &   2.973   &   2.975  \\
001245.12$-$054945.7   &   2.214   &   2.202  \\
001316.82$-$093841.2   &   2.644   &   2.628  \\
002144.10+172601.1   &   2.075   &   2.078  \\
003606.10+272539.2   &   2.180   &   2.159  \\
003901.47+073434.2   &   2.267   &   2.274  \\
00473.084+140228.8   &   2.178   &   2.159  \\
005321.17+112240.6   &   2.595   &   2.575  \\
\hline
\end{tabular}
\label{newghostly}
\end{table}

\begin{figure*}
\centering
\begin{tabular}{c}
\includegraphics[width=0.95\hsize]{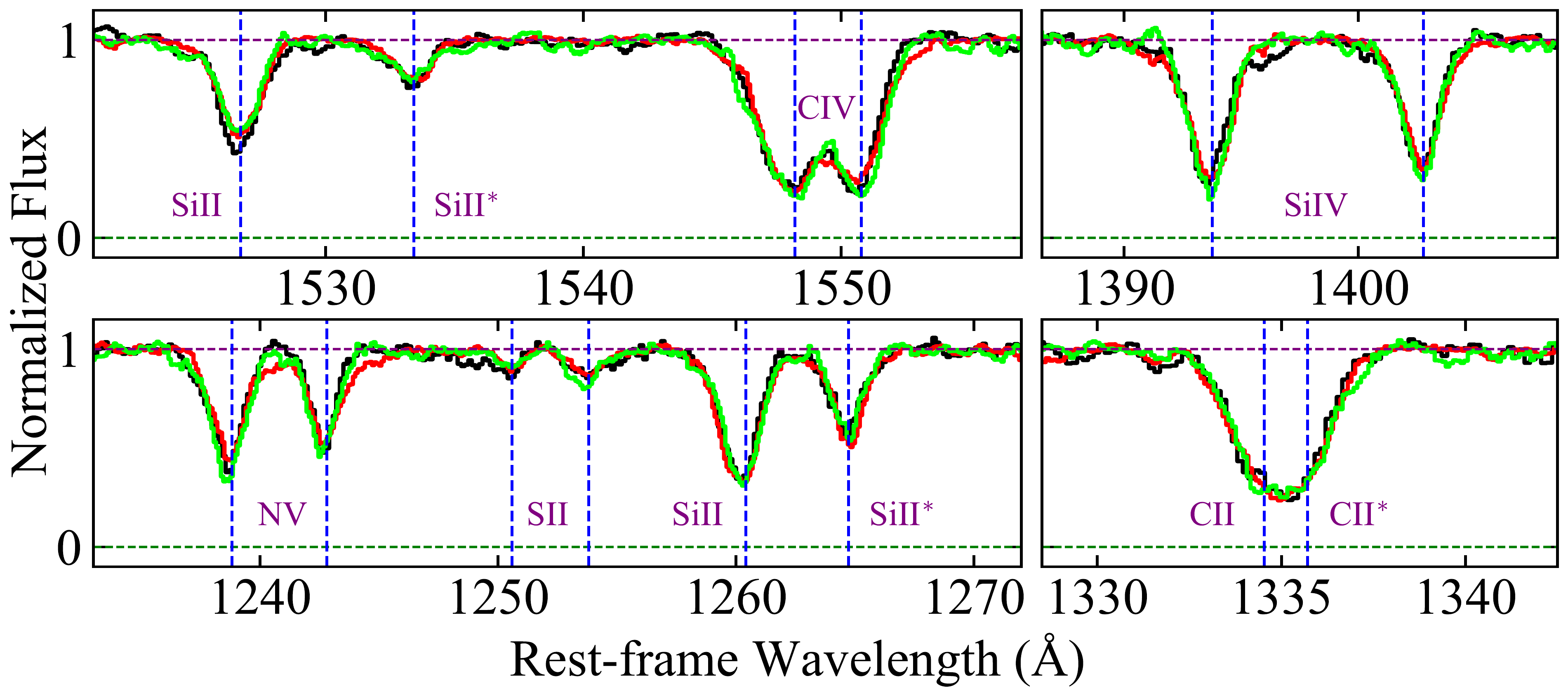}
\end{tabular}
\caption{Some important transitions detected in the stacked spectrum of \citetalias{2020ApJ...888...85F} (black spectrum) and of the current work (red and green spectra). Here, the green lines are from the stacked spectrum of 21 ghostly absorbers with strong Ly$\beta$ absorption, and the red lines are from the stacked spectrum of the 68 remaining spectra.}
 \label{metals}
\end{figure*}

\subsection{Constraining the {\rm H}\,{\sc i} Column Density} \label{sec:HI_column_density}

The absence of the Ly$\alpha$ absorption in the spectra of ghostly absorbers makes it difficult to robustly determine the H\,{\sc i} column density of these absorbers, and consequently confirm whether they are indeed DLAs. In \citetalias{2020ApJ...888...85F}, we employed three methods to infer whether a ghostly absorber is a DLA: 1) directly measuring the H\,{\sc i} column density of the absorber using absorption from other Lyman series lines, 2) estimating the H\,{\sc i} column density using the shallow dip seen around the Ly$\alpha$ spectral region, and 3) comparing the EWs of some strong metal absorption lines detected in the spectra of ghostly absorbers with those detected in intervening DLAs.

Out of these three approaches, the first approach is the most robust one as it would allow us to directly measure the H\,{\sc i} column density. However, this approach requires at least the Ly$\beta$ absorption to exhibit the characteristic damping wings. If the damping wings of the Ly$\beta$ absorption are not strong enough, it will be difficult to detect them in low resolution spectra especially when the data is noisy. Unfortunately, the Ly$\beta$ spectral region in most of our spectra has low SNR. This limits our ability to accurately measure the H\,{\sc i} column density in individual systems. We, therefore, try to stack these spectra to increase the SNR. The higher SNR of the stacked spectrum would in principle allow us to determine the H\,{\sc i} column density by fitting the damping wings of the Ly$\beta$ absorption. If the majority of the systems have high H\,{\sc i} column density, the damping wings of the Ly$\beta$ absorption should be clearly and prominently visible in the stacked spectrum. In order to increase the chances of clearly detecting the damping wings of the Ly$\beta$ absorption in the stacked spectrum, we carefully choose the quasar spectra that are used in the construction of the stacked spectrum using the following procedure. 

We first create a synthetic Ly$\beta$ absorption line with H\,{\sc i} column density of $N$(H\,{\sc i})\,=\,10$^{21}$\,cm\,$^{-2}$ and $b$-value of 100\,km\,s$^{-1}$. We then exclude quasar spectra whose Ly$\beta$ absorption are narrower than this synthetic absorption line. For example, in the lower panel (resp. upper panel) of Fig.\,\ref{goodbad}, we show two quasar spectra with their Ly$\beta$ absorption being narrower (resp. broader) than the synthetic one. This conservative approach ensures that only spectra of the absorbers with relatively higher H\,{\sc i} column density and stronger damping wings contribute in the construction of the stacked spectrum. Applying this constraint on the $S_{\beta}$ sample leaves us with 21 spectra. We then median-stack these spectra following the method described in \citetalias{2020ApJ...888...85F}.

Figure\,\ref{lyabg} shows the resulting stacked spectrum around the Ly$\alpha$, Ly$\beta$, and Ly$\gamma$ spectral regions. The Voigt profile fit for log\,$N$(H\,{\sc i})\,=\,21.5 and $b$-value of 50\,km\,s$^{-1}$ is overplotted on the observed spectrum as the red curve. This column density is 0.5\,dex higher than the H\,{\sc i} column density derived from the stacked spectrum of \citetalias{2020ApJ...888...85F}. As shown in Fig.\,\ref{lyabg}, the damping wings (especially the red wing) of the Ly$\beta$ absorption is clearly visible in the spectrum. This implies that the majority of the ghostly absorbers whose spectra are used in the stacked spectrum, are DLAs. Follow-up observation with higher resolution and SNR are needed to precisely measure the H\,{\sc i} column density in the individual systems, and confirm which one of them are indeed DLAs.

\subsection{Metal Absorption Lines in Ghostly DLAs}

In \citetalias{2020ApJ...888...85F}, we created a stacked spectrum using the spectra of 30 ghostly absorbers found in the SDSS DR12. By comparing the metal absorption lines detected in the stacked spectra of ghostly absorbers and eclipsing DLAs, we found that ghostly absorbers have stronger absorption from both high ioniztion species like N\,{\sc v}, Si\,{\sc iv}, and C\,{\sc iv} and fine structure states of Si\,{\sc ii}, C\,{\sc ii}, and O\,{\sc i}. Due to the lack of information on the H\,{\sc i} column density of most of these ghostly absorbers, there was a concern that non-DLA absorbers such as SLLSs could have important contribution in the properties of the absorption lines in the stacked spectrum. However, thanks to the higher number of ghostly absorbers with Ly$\beta$ absorption in the current sample, we decreased this concern by stacking only the spectra of the absorbers that exhibit apparently broad Ly$\beta$ absorption, which is indicative of a high H\,{\sc i} column density. 

Figure\,\ref{metals} shows some metal absorption lines detected in the stacked spectrum of \citetalias{2020ApJ...888...85F} (black spectrum) and of the current work (red and green spectra). Here, the green lines are from the stacked spectrum of 21 ghostly absorbers with broad Ly$\beta$ absorption (see section\,\ref{sec:HI_column_density}), and the red lines are from the stacked spectrum of the 68 remaining spectra. These remaining spectra either do not cover the Ly$\beta$ spectral region or do not exhibit broad enough Ly$\beta$ absorption. The fact that the black and red absorption lines are consistent with the green lines implies that the majority of the ghostly absorbers from \citetalias{2020ApJ...888...85F} and from this work are highly likely DLAs.Therefore, the results from \citetalias{2020ApJ...888...85F} could be generalized to ghostly DLAs.

\section{Summary and Conclusions}

In this paper, we applied machine learning on the SDSS DR14 spectra to find candidate DLA absorbers with $z_{\rm abs}$\,$\sim$\,$z_{\rm QSO}$ using metal absorption line properties as the input features to the model. We then inspected these candidate DLAs and found 89 ghostly absorbers of which 59 are new systems. The spectra of 41 ghostly absorbers also cover the Ly$\beta$ spectral region. We constructed a median-stacked spectrum using the spectra of 21 ghostly absorbers which exhibit strong Ly$\beta$ absorption. By fitting the damping wings of the Ly$\beta$ absorption in the stacked spectrum, we measured an H\,{\sc i} column density of log\,$N$(H\,{\sc i})\,=\,21.50. This is 0.5 dex larger than the H\,{\sc i} column density measured from the stacked spectrum of \citetalias{2020ApJ...888...85F}. The high H\,{\sc i} column density measured in this work strongly suggests that the majority of the absorbers used in the construction of the stacked spectrum are indeed DLAs. The metal absorption lines in this stacked spectrum are also consistent with those of \citetalias{2020ApJ...888...85F}.

In \citetalias{2020ApJ...888...85F}, by analyzing the metal absorption lines of ghostly absorbers, we proposed a scenario in which ghostly absorbers, compared to eclipsing DLAs, are denser and located closer to the central engine of the quasar. The similarity that we found between the metal absorption lines of this work and those of \citetalias{2020ApJ...888...85F} suggests that this scenario could also be applied to ghostly DLAs. 

Although the circumstantial evidence presented in this work strongly suggest that the majority of our ghostly absorbers are indeed DLAs, follow-up high resolution and high SNR spectroscopic observation of individual systems are needed to robustly confirm this.

\section*{Acknowledgements}
\noindent
We would like to thank the referee for the useful comments which improved the quality of the paper.
Funding for SDSS-III has been provided by the Alfred P. Sloan Foundation, the Participating Institutions, the National Science Foundation, and the U.S. Department of Energy Office of Science. The SDSS-III web site is \url{http://www.sdss3.org/}.
\noindent
SDSS-III is managed by the Astrophysical Research Consortium for the Participating Institutions of the SDSS-III Collaboration including the University of Arizona, the Brazilian Participation Group, Brookhaven National Laboratory, Carnegie Mellon University, University of Florida, the French Participation Group, the German Participation Group, Harvard University, the Instituto de Astrofisica de Canarias, the Michigan State/Notre Dame/JINA Participation Group, Johns Hopkins University, Lawrence Berkeley National Laboratory, Max Planck Institute for Astrophysics, Max Planck Institute for Extraterrestrial Physics, New Mexico State University, New York University, Ohio State University, Pennsylvania State University, University of Portsmouth, Princeton University, the Spanish Participation Group, University of Tokyo, University of Utah, Vanderbilt University, University of Virginia, University of Washington, and Yale University.

\bibliographystyle{aasjournal}
\bibliography{ref}

\end{document}